\documentclass[onecolumn]{IEEEtran}
\usepackage{amsmath,amsfonts,amssymb} 
\usepackage{amsthm} 
\usepackage[utf8]{inputenc}
\usepackage{algorithmic}
\usepackage{titlesec}
\usepackage{algorithm}
\usepackage{array}
\usepackage{textcomp}
\usepackage{stfloats}
\usepackage{url}
\usepackage{verbatim}
\usepackage{graphicx}
\usepackage{cite}
\usepackage{amsthm}%
\usepackage{mathrsfs}%
\usepackage{xcolor}%
\usepackage{textcomp}%
\usepackage{manyfoot}
\usepackage{array,booktabs}%
\usepackage{syntax}
\usepackage{fancybox}
\usepackage{adjustbox}
\usepackage{physics}
\usepackage{soul}
\usepackage{listings}
\usepackage{multirow}
\usepackage{multicol}
\usepackage{comment}
\usepackage{hyperref}
\usepackage{nameref}
\usepackage{mathtools}
\usepackage{tikz-cd}
\usepackage{float}
\usepackage{titlesec}
\titleformat{\subparagraph}[runin]{\normalfont\normalsize\bfseries}{\thesubparagraph}{1em}{}
\usepackage{tocloft}
\newtheorem{theorem}{Theorem}
\newtheorem{lemma}{Lemma}
\newtheorem{proposition}{Proposition}
\newtheorem{corollary}{Corollary}
\newtheorem{definition}{Definition}
\newtheorem{construction}{Construction}
\newtheorem{example}{Example}
\newtheorem{remark}{Remark}

\newcommand{\F}{\mathbb{F}}
\newcommand{\Fq}{\mathbb{F}_q}
\newcommand{\C}{\mathbb{C}}
\newcommand{\cS}{\mathcal{S}}
\newcommand{\cB}{\mathcal{B}}
\newcommand{\cC}{\mathcal{C}}
\newcommand{\cL}{\mathcal{L}}
\newcommand{\cP}{\mathcal{P}}

\newcommand{\rk}{\operatorname{rank}}
\newcommand{\Span}{\operatorname{span}}
\newcommand{\row}{\operatorname{row}}
\newcommand{\supp}{\operatorname{supp}}
\newcommand{\wt}{\operatorname{wt}}
\newcommand{\cl}{\operatorname{cl}}
\newcommand{\cutrk}{\operatorname{cutrk}}
\newcommand{\GraphS}[1]{\mathcal{S}(#1)}
\newcommand{\nuo}{\nu_{\circ}}
\newcommand{\bd}{\partial}
\newcommand{\Fz}[1]{\F^{#1}_{0}}
\newcommand{\one}{\mathbf{1}}
\newcommand{\e}{\mathbf{e}}
\newcommand{\Dh}{\widehat{D}}
\newcommand{\Grec}{\Gamma^{\mathrm{rec}}}
\newcommand{\Gforb}{\Gamma^{\mathrm{forb}}}
\newcommand{\Gint}{\Gamma^{\mathrm{int}}}
\newcommand{\Gcirc}{\Gamma^{\mathrm{circ}}}
\newcommand{\Gmin}{\Gamma^{\mathrm{min}}}

\begin{document}

\title{Symplectic Matroids, Stabilizer Codes, and Quantum Secret-Sharing}
\author{
    Sayantan Ganguly\(^{1,2}\)\\
    \(^{1}\)Institute for Advancing Intelligence (IAI), TCG CREST, Kolkata, India\\
    \(^{2}\)Ramakrishna Mission Vivekananda Educational and Research Institute, Belur\\
}


\maketitle
\begin{abstract}
An $\mathbb{F}_q$-linear $[[n,k]]_q$ stabilizer code is represented by a totally isotropic subspace of $\mathbb{F}_q^{2n}$ of dimension $n-k$, whose full-rank generator matrix has nonzero maximal minors on admissible column sets forming the bases of a representable symplectic matroid. Sarvepalli asked whether graphical symplectic matroids introduced by Chow give stabilizer codes with good parameters, when a Lagrangian matroid induces a quantum access structure, which code realizes it, and which polynomial carries the weight enumerator. We answer these questions. For a $3$-connected simple graph, every representation of a graphical symplectic matroid, over any field, gives a stabilizer code of distance exactly $2$, so the girth is not a lower bound. For a represented Lagrangian matroid and a dealer, we construct the dealer-shortened stabilizer code, its logical operators, an encoding isometry and explicit recovery maps, and we show that this is the code implemented by the stabilizer state through teleportation. A set of players recovers the secret, learns nothing, or learns part of it according to a rank formula on pairs of columns, which equals its mutual information with the dealer; for graph states it becomes a cut-rank condition. Over $\F_2$ a symplectic matroid of any rank has only one representation, so the code, its access structure and its weight enumerators depend on the matroid alone; for $q\ge3$ even the enumerator does not. Circuit-induced authorized sets always reconstruct when the matroid is self-dual, but not in general, and a six-qubit graph state shows that the two access structures can differ even then. Finally, a polynomial in the ranks of column pairs determines both weight enumerators.
\end{abstract}

\begin{IEEEkeywords}
Symplectic matroids, stabilizer codes, quantum secret-sharing, weight enumerators.
\end{IEEEkeywords}

\section*{Contents}
\tableofcontents

\section{Introduction}\label{sec:intro}

\IEEEPARstart{A}{n} $\Fq$-linear stabilizer code on $n$ qudits is described by an isotropic subspace of $\Fq^{2n}$ with respect to the standard alternating form \cite{CRSS1998,KetkarEtAl2006}. If $H=[X\mid Z]$ is a generator matrix of this subspace, with the columns indexed by $J=[n]\sqcup[n]^*$, then the column
sets that contain at most one of the two columns $i,i^*$ of every qudit and carry a nonzero maximal minor are the bases of a \emph{symplectic matroid}, a Coxeter matroid of type $C_n$
\cite{BorovikGelfandWhite2003}. Sarvepalli \cite{Sarvepalli2011} made this connection explicit:
an $[[n,n-r]]_q$ code gives a representable symplectic matroid of rank $r$, and conversely. At the end of his paper he asked the following questions \cite[Sec.~V]{Sarvepalli2011}.
\begin{enumerate}
\item Find representations of Chow's graphical symplectic matroids \cite{Chow2003}.
\item Do the quantum codes derived from the symplectic matroids of simple connected graphs have good
parameters?
\item Find necessary and sufficient conditions for a Lagrangian matroid to induce a quantum access structure, if possible, in terms of the underlying graph.
\item Given a secret-sharing Lagrangian matroid, which quantum code realizes its access structure?
\item Find a polynomial that captures the weight enumerator of the underlying code.
\end{enumerate}
Question~1 is treated in the paper \cite{GangulyPaperI}. Here we answer Questions~2--5. From \cite{GangulyPaperI} we use only the statements recalled in Theorem~\ref{thm:paperI}, and only for Question~2. In Appendix~\ref{app:envelopes}, we give the ordinary-envelope constructions and, based on matroid operations, we construct a family of ordinary envelopes of symplectic matroids which are relevant to coding invariants. In a separate companion paper~\cite{GangulyPaperI}, we have addressed the representability theory of graphical symplectic matroids.

In a representable symplectic matroid, we can define the ranks of \emph{admissible} column sets, those containing at most one column of each qudit. The questions that matter for codes, namely which operators are supported on a set of qudits and which sets of players can recover a secret, are governed instead by the ranks of the full column pairs $\{i,i^*\}$, $i\in A$. We call these the \emph{pair ranks} $\rho(A)$. Much of what follows is about the gap between the two kinds of rank: when the pair ranks are forced by the matroid, and what goes wrong when they are not.

\subsection{Main results}

\paragraph*{Stabilizer codes from graphs (Question~2)} Let $G=(V,E)$ be a connected graph, and contains a cycle. If graphical symplectic matroid $\GraphS{G}$ is representable over $\Fq$, every representation gives an $[[|E|,|E|-|V|]]_q$ code. When $|E|>|V|$, and if it has distance $d$, it was suggested in \cite[Sec.~III]{Sarvepalli2011} that $d$ is at least the girth of $G$. We show that $d\ge2$ when $G$ has no loops and no bridges (Theorem~\ref{thm:weight-one}), and that $d\le2$ for every representation over every finite field as soon as $G$ contains a pair of edges in one of two configurations (Theorem~\ref{thm:barrier}). Hence every $3$-connected simple graph gives distance exactly~$2$ (Corollary~\ref{cor:3conn}); together with the classification in \cite{GangulyPaperI} this proves the $3$-connected case completely (Remark~\ref{rem:Q2}). The girth bound fails already for $K_4$, where we give the code explicitly (Example~\ref{ex:K4}), while a subdivided graph, the theta graph $K_{2,3}$, has a $[[6,1,3]]_4$ representation (Example~\ref{ex:theta}). We prove the contraction and truncation on symplectic matroids and derive associated stabilizer codes (Propositions~\ref{prop:contraction} and~\ref{prop:truncation}) and distinguish stabilizer extension from ordinary Higgs lifts (Remark~\ref{rem:higgs}, Appendix~\ref{app:envelopes}).

\paragraph*{The code of a secret-sharing scheme (Questions~3 and~4)} Fix a Lagrangian matroid $L$
with a representation $W\le\Fq^{2n}$ and a dealer $d$. Shortening $W$ at the dealer gives a stabilizer
code $Q_d(W)$ on the $n-1$ players, which encodes one qudit exactly when the dealer projection is
onto. We construct its stabilizer matrix and logical operators directly from a generator matrix of
$W$ (Theorem~\ref{thm:Hd}, Construction~\ref{con:dealer}), identify the dealer coordinates with its
logical space by an anti-symplectic isomorphism (Theorem~\ref{thm:Theta}), and give an encoding
isometry and explicit recovery maps (Theorem~\ref{thm:encoding}). The stabilizer state of $W$ is the Choi state of an encoding onto this code, so teleportation through the stabilizer state realizes the scheme
(Proposition~\ref{prop:choi}). We compute its access structure (Theorem~\ref{thm:operational}): a player set $A$ recovers the secret, learns nothing, or learns part of it according as $\rho(\bar A\cup\{d\})-\rho(\bar A)$ equals $2$, $0$ or $1$, and this number is the mutual information $I(A:d)$ in units of $\log q$. For graph states, it is a cut-rank condition (Corollary~\ref{cor:cutrank}). Intermediate sets can be removed. It is possible to construct a mixed perfect hybrid secret-sharing scheme (Proposition~\ref{prop:hybrid}), and the CSS construction of Sarvepalli and Raussendorf is the special case of a code and its dual (Proposition~\ref{prop:css}).

\paragraph*{Uniqueness of representability over \(\mathbb{F}_{2}\)} Every $\F_2$-representable symplectic matroid, of any rank, has exactly
one binary isotropic row space (Theorem~\ref{thm:rigidity}). Over $\F_2$ the code up to Pauli conjugation, its access structures
and its weight enumerators are therefore invariants of the matroid. For every $q\ge3$ this fails, even
for the weight enumerator (Corollary~\ref{cor:dichotomy}).

\paragraph*{Circuit-induced versus operational access structures} Following Sarvepalli and
Raussendorf \cite{SarvepalliRaussendorf2010}, Sarvepalli attaches to $L$ and $d$ a circuit-induced
access structure $\Gcirc_{L,d}$ \cite{Sarvepalli2011}. We characterize when it is a quantum access
structure (Proposition~\ref{prop:nocloning}), prove through a circuit-cocircuit orthogonality that
self-dual representable matroids satisfy the intersection condition, with circuit nonemptiness required separately (Theorem~\ref{thm:selfdual-ss}), describe it through vectors of $W$ with admissible
support (Proposition~\ref{prop:admissible-lift}), and obtain the necessary and sufficient conditions for circuit-induced access structures to be a quantum access-structure and consequently the necessary and sufficient condition for a Lagrangian matroid to induce a quantum secret-sharing scheme (Theorem~\ref{thm:graph-circuits}). If $L$ is self-dual, every circuit-induced authorized set recovers the secret (Theorem~\ref{thm:selfdual}); without self-duality this fails already for three qudits (Example~\ref{ex:looped}). Theorem~\ref{thm:comparison} states exactly what is needed for
equality and for a perfect realization. Over $\F_2$, the graph states with a self-dual matroid are those with $A^2=I$, which includes every graph whose vertices come in adjacent twin pairs
(Proposition~\ref{prop:graph-selfdual}). The smallest such example with three pairs is a six-qubit graph state that is secret-sharing for every dealer, yet its operational access structure is strictly larger than the circuit-induced one for every dealer (Theorem~\ref{thm:Gstar}); among graph states of
simple graphs whose circuit-induced family at the chosen dealer is a nonempty quantum access structure, six qubits is the smallest size at which the two structures differ (Table~\ref{tab:census}).

\paragraph*{Enumerators (Question~5)} We construct polynomial from the pair ranks, which determines the stabilizer and normalizer weight enumerators, hence the distance and the MacWilliams identity (Theorem~\ref{thm:enumerator}). It is a matroid invariant when $q=2$ and not when $q\ge3$ (Corollary~\ref{cor:enum-invariance}).

\subsection{Related work}

Quantum secret-sharing was introduced by Hillery, Bu\v{z}ek and Berthiaume \cite{HilleryBuzekBerthiaume1999}
and by Cleve, Gottesman and Lo \cite{CleveGottesmanLo1999}; general access structures were treated by
Gottesman \cite{Gottesman2000} and Smith \cite{Smith2000}, and an information-theoretic formulation is
due to Imai, M\"uller-Quade, Nascimento, Tuyls and Winter \cite{ImaiEtAl2005}. Schemes based on graph
states were developed by Markham and Sanders \cite{MarkhamSanders2008}, Sarvepalli
\cite{Sarvepalli2012}, Javelle, Mhalla and Perdrix \cite{JavelleMhallaPerdrix2012}, and Marin and
Markham \cite{MarinMarkham2013}. Matroids entered the subject through identically self-dual matroids
\cite{SarvepalliRaussendorf2010} and Lagrangian matroids \cite{Sarvepalli2011}. Lagrangian matroids
are Bouchet's delta-matroids \cite{Bouchet1987,BorovikGelfandWhite2003}, and binary delta-matroids
are studied in \cite{BouchetDuchamp1991}. Codes from graphs in the stabilizer formalism go back to
Schlingemann and Werner \cite{SchlingemannWerner2001}. Greene's theorem \cite{Greene1976} relates
weight enumerators and Tutte polynomials; versions for $2$-polymatroids and additive codes are in
\cite{OxleyWhittle1993,SwartzWentworthNiceXue2024}, and the quantum MacWilliams identities are due to
Shor and Laflamme \cite{ShorLaflamme1997} and Rains \cite{Rains1999}.

Supported logical operators and their information groups were studied by Gheorghiu, Looi and Griffiths \cite{GheorghiuLooiGriffiths2010}. Gheorghiu \cite{Gheorghiu2012} determines stabilizer access structures and removes intermediate information by twirling and classical key sharing. Our supported-space formulation and hybrid construction use this established mechanism; the contributions here are the pair-rank expressions, their relation to admissible circuits, and the representation-dependence and rigidity results. The maximal-rank binary uniqueness theorem is also known \cite{Zhukov2018}; Theorem~\ref{thm:rigidity} treats every isotropic rank.

\subsection{Organization}
In Section~\ref{sec:prelim}, we fix notations and recall the facts on stabilizer codes and symplectic matroids that we use throughout the paper. In Section~\ref{sec:graph} we derive and discuss codes from graphical symplectic matroids; in Section~\ref{sec:qss} we discuss quantum secret-sharing, and in Section~\ref{sec:enum} we discuss enumerators. Lastly, in Section~\ref{sec:concl} we discuss open problems. In the appendices, we provide Python programs for computations, and we discuss the relation with ordinary envelopes and its limits for code construction.

\section{Preliminaries}\label{sec:prelim}

Throughout, $q=p^m$ with $p$ prime and $[n]=\{1,\dots,n\}$. Let $\sigma:\Fq\to\Fq$, $\sigma(x)=x^p$, be the
Frobenius automorphism; it generates the Galois group of $\Fq$ over $\F_p$, which has order $m$. The
\emph{absolute trace} is
\[ \tr_{\mathbb{F}_{q}/\mathbb{F}_{p}}(x)=\sum_{j=0}^{m-1}\sigma^j(x)=x+x^p+\dots+x^{p^{m-1}},\qquad x\in\Fq. \]
We use three standard facts \cite[Sec.~2.3]{LidlNiederreiter1997}. First, $\tr_{\mathbb{F}_{q}/\mathbb{F}_{p}}(x)\in\F_p$,
since $\tr_{\mathbb{F}_{q}/\mathbb{F}_{p}}(x)^p=\sum_{j=1}^{m}x^{p^j}=\tr_{\mathbb{F}_{q}/\mathbb{F}_{p}}(x)$ because $x^{p^m}=x$. Second, $\tr_{\mathbb{F}_{q}/\mathbb{F}_{p}}$ is $\F_p$-linear,
because $\sigma$ is additive and fixes $\F_p$, and it maps $\Fq$ onto $\F_p$, because as a polynomial of
degree $p^{m-1}<q$ it is not identically zero on $\Fq$. Third, the trace form $(x,y)\mapsto\tr_{\mathbb{F}_{q}/\mathbb{F}_{p}}(xy)$ is a nondegenerate $\F_p$-bilinear form on $\Fq$: if $x\neq0$, choose $z$ with $\tr_{\mathbb{F}_{q}/\mathbb{F}_{p}}(z)\neq0$, and then $y=x^{-1}z$ has $\tr_{\mathbb{F}_{q}/\mathbb{F}_{p}}(xy)\neq0$. Consequently every basis $\beta_1,\dots,\beta_m$ of $\Fq$ over $\F_p$ has a unique \emph{trace-dual} basis $\beta'_1,\dots,\beta'_m$, with $\tr_{\mathbb{F}_{q}/\mathbb{F}_{p}}(\beta_i\beta'_j)=\delta_{ij}$. If $q=p$,
then $\tr_{\mathbb{F}_{q}/\mathbb{F}_{p}}$ is the identity map of $\F_p$. Complements of sets of qudits are always taken in a set that is clear from the context; it is $[n]$ unless stated otherwise.

All subspaces of $\Fq^N$ that occur are $\Fq$-linear unless we say otherwise, and for such a subspace
$U$ we write $\dim_{\Fq}U$ for its dimension over $\Fq$. For a field $\F$ and a set $X$ of vectors over
$\F$, $\Span_{\F}X$ is the $\F$-linear span of $X$, and for a matrix $M$ over $\Fq$, $\row(M)$ is the
$\Fq$-span of its rows. The dimension of a Hilbert space is its complex dimension, written $\dim_{\C}$.
For $\beta\in\Fq$, $-\beta$ is the additive inverse of $\beta$ in $\Fq$; in characteristic $2$,
$-\beta=\beta$.

\subsection{The symplectic space}\label{subsec:space}

We index the coordinates of $\Fq^{2n}$ by $J=[n]\sqcup[n]^*$, where $[n]^*=\{i^*:i\in[n]\}$, and write
a vector as $v=(a\mid b)$ with $a,b\in\Fq^n$, so that $v_i=a_i$ and $v_{i^*}=b_i$. The symplectic form
is
\begin{equation}\label{eq:omega}
 \omega\big((a\mid b),(a'\mid b')\big)=a\cdot b'-b\cdot a' = v\,\Omega_n\,v'^{T},
\end{equation}
where $v=(a\mid b)$, $v'=(a'\mid b')$ and
$\Omega_n=\left(\begin{smallmatrix}0&I_n\\-I_n&0\end{smallmatrix}\right)$.
For a subspace $U$ we write $U^\perp$ for its orthogonal complement with respect to $\omega$; $U$ is
\emph{isotropic} if $U\subseteq U^\perp$ and \emph{Lagrangian} if moreover $\dim_{\Fq} U=n$.

The two coordinates $i,i^*$ belong to qudit $i$. For $A\subseteq[n]$ put $J_A=A\sqcup A^*$ and let
$E_A\le\Fq^{2n}$ be the space of vectors supported on $J_A$. The form restricted to $E_{\{i\}}$ has
Gram matrix $\left(\begin{smallmatrix}0&1\\-1&0\end{smallmatrix}\right)$ in every characteristic,
and vectors with disjoint supports are orthogonal. Hence each $E_A$ is a nondegenerate symplectic
space and
\begin{equation}\label{eq:split}
 \Fq^{2n}=E_A\perp E_{\bar A},\qquad E_A^\perp=E_{\bar A}.
\end{equation}
More generally, if $A,B\subseteq[n]$ are disjoint, then $E_A\cap E_B=0$ and $E_A+E_B=E_{A\cup B}$, and
$E_A$ and $E_B$ are orthogonal. We write $E_A\oplus E_B=E_{A\cup B}$ for this internal orthogonal direct
sum, and similarly $U\oplus U'=U+U'$ for subspaces $U\le E_A$ and $U'\le E_B$. Every vector of
$E_A\oplus E_B$ is uniquely $u+u'$ with $u\in E_A$, $u'\in E_B$; we also write it as the pair $(u,u')$,
and then $\omega\big((u,u'),(w,w')\big)=\omega(u,w)+\omega(u',w')$.

For $v\in\Fq^{2n}$ the \emph{support} $\supp(v)\subseteq[n]$ is the set of qudits $i$ with $(v_i,v_{i^*})\neq(0,0)$, the \emph{weight} is $\wt(v)=|\supp(v)|$, and $\supp_J(v)\subseteq J$ is the set of coordinates $j$ with $v_j\neq0$.

The form $\omega$ is \emph{alternating}, that is, $\omega(v,v)=0$ for every $v$; this matters in
characteristic $2$. We shall use the following standard fact.

\begin{lemma}[Symplectic bases]\label{lem:symplectic-basis}
Let $U$ be a vector space of dimension $2m$ over a field $\F$ with a nondegenerate alternating form $\omega$, and let $u,v\in U$ with $\omega(u,v)=1$. Then $U$ has a basis
$u_1,v_1,\dots,u_m,v_m$ with $u_1=u$, $v_1=v$, $\omega(u_i,v_j)=\delta_{ij}$ and $\omega(u_i,u_j)=\omega(v_i,v_j)=0$ for all $i,j$. Consequently, if also $u',v'\in U$ satisfy
$\omega(u',v')=1$, there is an $\F$-linear bijection $F \colon U\to U$ with $\omega(Fx,Fy)=\omega(x,y)$ for all
$x,y$, $F(u)=u'$ and $F(v)=v'$.
\end{lemma}
\begin{IEEEproof}
Let $P=\Span_{\F}\{u,v\}$. The Gram matrix on $P$ is
$\left(\begin{smallmatrix}0&1\\-1&0\end{smallmatrix}\right)$, and
$x-\omega(x,v)u+\omega(x,u)v\in P^\perp$ for every $x\in U$.
Thus $U=P\oplus P^\perp$. The restricted form on $P^\perp$ is nondegenerate: a vector in its
radical is orthogonal to both summands, hence to $U$.
Inductively choose $u_2\ne0$ in $P^\perp$ and $v_2$ with $\omega(u_2,v_2)=1$, and repeat in
their orthogonal complement. This gives the required basis. Mapping a resulting basis
extending $(u,v)$ to one extending $(u',v')$ preserves all pairings and proves the last assertion.
\end{IEEEproof}

\subsection{Stabilizer codes}\label{subsec:stabilizer}

Let $\chi(x)=e^{2\pi i\,\tr_{\mathbb{F}_{q}/\mathbb{F}_{p}}(x)/p}$ for $x\in\Fq$, where $\tr_{\mathbb{F}_{q}/\mathbb{F}_{p}}(x)\in\F_p$ is identified with an integer in $\{0,\dots,p-1\}$. Then $\chi(x+y)=\chi(x)\chi(y)$, because $\tr_{\mathbb{F}_{q}/\mathbb{F}_{p}}$ is additive and $e^{2\pi ik/p}$ depends only on $k$ modulo $p$; and $\chi$ is not identically $1$, because $\tr_{\mathbb{F}_{q}/\mathbb{F}_{p}}$ is onto. Thus $\chi$ is a nontrivial additive character of $\Fq$, and $\chi(cx)=1$ for all $c\in\Fq$ only if $x=0$, by the third fact above. On $\C^q$, with orthonormal basis $\{|x\rangle:x\in\Fq\}$, put $X(a)|x\rangle=|x+a\rangle$ and $Z(b)|x\rangle=\chi(bx)|x\rangle$; for $v=(a\mid b)\in\Fq^{2n}$ let $D(v)=X(a_1)Z(b_1)\otimes\dots\otimes X(a_n)Z(b_n)$. A direct computation gives
\begin{equation}\label{eq:commutation}
 D(u)D(v)=\chi\big(\omega(v,u)\big)\,D(v)D(u),
\end{equation}
so $D(u)$ and $D(v)$ commute if and only if $\tr_{\mathbb{F}_{q}/\mathbb{F}_{p}}(\omega(u,v))=0$. The operators $D(v)$ form a basis of
the operators on $(\C^q)^{\otimes n}$, and $\Tr D(v)=0$ for $v\neq0$. Let $\mu$ be the group of
$p$-th roots of unity if $p$ is odd and the group $\{\pm1,\pm i\}$ if $p=2$; the operators $cD(v)$ with
$c\in\mu$ and $v\in\Fq^{2n}$ are the \emph{Pauli operators}.

\begin{lemma}\label{lem:trace}
If $U\le\Fq^{2n}$ is an $\Fq$-subspace, then $\{v\colon\tr_{\mathbb{F}_{q}/\mathbb{F}_{p}}(\omega(u,v))=0\ \forall u\in U\}=U^\perp$. In particular, $U$ is isotropic for $\omega$ if and only if it is isotropic for $\tr_{\mathbb{F}_{q}/\mathbb{F}_{p}}\circ\omega$.
\end{lemma}
\begin{IEEEproof}
If $\tr_{\mathbb{F}_{q}/\mathbb{F}_{p}}(\omega(u,v))=0$ for all $u\in U$, then $\tr_{\mathbb{F}_{q}/\mathbb{F}_{p}}(c\,\omega(u,v))=\tr_{\mathbb{F}_{q}/\mathbb{F}_{p}}(\omega(cu,v))=0$ for all $c\in\Fq$, and nondegeneracy of the trace form gives $\omega(u,v)=0$.
\end{IEEEproof}

Here we need coherent phases to make the restriction of $D$ to an isotropic subspace a homomorphism. In odd characteristic one may take $\Dh(a\mid b)=\chi(\tfrac12a\cdot b)D(a\mid b)$ on that subspace. In characteristic two, $D(v)^2=\chi(a\cdot b)I$ may equal $-I$, so fourth roots of unity are needed, as for $XZ$.

\begin{lemma}[Stabilizer lifts]\label{lem:lift}
Let $S\le\Fq^{2n}$ be isotropic with $\dim_{\Fq}S=r$. There are phases $c(s)\in\mu$ such that
$\Dh(s)=c(s)D(s)$ satisfies $\Dh(s+s')=\Dh(s)\Dh(s')$ for all $s,s'\in S$. The operators $\Dh(s)$
form an abelian group of order $q^r$ whose only scalar element is $I$, and
\[ \Pi_S=q^{-r}\sum_{s\in S}\Dh(s) \]
is the orthogonal projection onto a subspace $Q$ of $(\C^q)^{\otimes n}$ with $\dim_{\C}Q=q^{n-r}$.
\end{lemma}
\begin{IEEEproof}
As an $\F_p$-vector space, $S$ has dimension $mr$; let $v_1,\dots,v_{mr}$ be a basis of $S$ over $\F_p$. For $v=(a\mid b)$ we have
$D(v)^p=\chi\big(\tbinom p2\,a\cdot b\big)I$, which is $I$ if $p$ is odd and $\pm I$ if $p=2$; choose
$c_j\in\mu$ with $(c_jD(v_j))^p=I$. The operators $c_jD(v_j)$ commute by \eqref{eq:commutation} and
Lemma~\ref{lem:trace}, so $\Dh(\sum_j\varepsilon_jv_j)=\prod_j(c_jD(v_j))^{\varepsilon_j}$,
$0\le\varepsilon_j<p$, is a well-defined homomorphism from $(S,+)$. By \eqref{eq:commutation},
$\Dh(s)$ is a multiple of $D(s)$ by an element of $\mu$. It is a scalar only if $s=0$, because the
$D(v)$ are linearly independent; in particular $s\mapsto\Dh(s)$ is injective, and the group has
order $|S|=q^r$. Hence $\Pi_S$ is $q^{-r}$ times the sum over a finite group of
unitaries, so it is an orthogonal projection, and its trace is $q^{-r}\Tr I=q^{n-r}$.
\end{IEEEproof}

The space $Q$ is the \emph{stabilizer code} of $S$ for the chosen lift. Two lifts differ by an additive character of $S$. Nondegeneracy of $\tr_{\mathbb{F}_{q}/\mathbb{F}_{p}}\circ\omega$ implies that
every such character is the restriction of $s\mapsto\chi(\omega(s,v))$ for some $v$.
Pauli conjugation therefore relates the two codespaces; dimensions, distances, supported logical
spaces, access families and weight enumerators are independent of the lift. We call $Q$ an $[[n,k]]_q$ code, $k=n-r$, and any generator matrix of $S$ a \emph{stabilizer matrix}. For $v\in S^\perp$ the operator $D(v)$ commutes with every $\Dh(s)$ and therefore maps $Q$ to itself; such operators are
\emph{logical}. If $v\in S$, then $D(v)$ acts on $Q$ as a scalar. The quotient \[ \cL(S)=S^\perp/S \]
is the space of \emph{logical classes}. The form $\omega$ induces a nondegenerate form on $\cL(S)$, because the radical of $\omega$ on $S^\perp$ is $S^\perp\cap S^{\perp\perp}=S$; and
$\dim_{\Fq}\cL(S)=(2n-r)-r=2k$. If $k\ge1$, the \emph{minimum distance} is
\begin{equation}\label{eq:distance}
 d(S)=\min\{\wt(v):v\in S^\perp\setminus S\},
\end{equation}
and $Q$ is an $[[n,k,d(S)]]_q$ code. The next lemma is the standard erasure criterion
\cite{KnillLaflamme1997,CleveGottesmanLo1999,Gottesman2000}; we include the short proof because the
argument is used again in Section~\ref{sec:qss}.

\begin{lemma}\label{lem:erasure}
Let $Q$ be the stabilizer code of an isotropic $S$ with $k\ge1$, and let $B\subseteq[n]$. The following
are equivalent.
\begin{enumerate}
\item The erasure of the qudits in $B$ can be corrected; that is, there is a channel acting on the
qudits in $\bar B$ that recovers every state of $Q$ from its restriction to $\bar B$.
\item The reduced state on $B$ of a state of $Q$ does not depend on the state.
\item $S^\perp\cap E_B\subseteq S$; that is, no nontrivial logical class is supported on $B$.
\end{enumerate}
\end{lemma}
\begin{IEEEproof}
By the Knill-Laflamme conditions \cite{KnillLaflamme1997}, (1) holds if and only if
$\Pi_SF\Pi_S=c(F)\Pi_S$ for every operator $F$ supported on $B$. Condition (2) says that
$\Tr(F\varrho)$ takes the same value on all states $\varrho$ of $Q$, for every such $F$; this is the
same condition, since an operator $T$ on $Q$ with $\langle\psi|T|\psi\rangle=0$ for all
$|\psi\rangle\in Q$, is zero. The operators $D(v)$, $v\in E_B$, span the operators supported on $B$.
If $v\notin S^\perp$, choose $s\in S$ and, using Lemma~\ref{lem:trace}, rescale it so that
$\zeta=\chi(\omega(s,v))\neq1$; then
$\Pi_SD(v)\Pi_S=\Pi_SD(v)\Dh(s)\Pi_S=\zeta\,\Pi_S\Dh(s)D(v)\Pi_S=\zeta\,\Pi_SD(v)\Pi_S$, so
$\Pi_SD(v)\Pi_S=0$. If $v\in S$, then $D(v)\Pi_S$ is a multiple of $\Pi_S$. Let
$v\in S^\perp\setminus S$. As the induced form on $\cL(S)$ is nondegenerate, there is $u\in S^\perp$
with $\omega(v,u)\neq0$, and after rescaling $u$ we may assume $\zeta=\chi(\omega(v,u))\neq1$. Both
$D(u)$ and $D(v)$ commute with $\Pi_S$. If $D(v)\Pi_S=\lambda\Pi_S$, then $\lambda\ne0$ because $D(v)$ is unitary on $Q$, and
$D(u)D(v)\Pi_S=\lambda D(u)\Pi_S=D(v)D(u)\Pi_S$, while \eqref{eq:commutation} gives
$D(u)D(v)=\zeta D(v)D(u)$; since $D(u)\Pi_S\neq0$, this forces $\zeta=1$, a contradiction. Hence
$\Pi_SD(v)\Pi_S$ is not a multiple of $\Pi_S$. Every operator $F$ supported on $B$ is a linear
combination of the $D(v)$, $v\in E_B$. By the three cases, $\Pi_SF\Pi_S$ is a multiple of $\Pi_S$ for
every such $F$ if no $v\in E_B$ lies in $S^\perp\setminus S$; and if some $v\in E_B$ does, then $F=D(v)$
violates the condition. This proves (1)$\Leftrightarrow$(3).
\end{IEEEproof}

For $A\subseteq[n]$ let $\cL_S(A)=\big((S^\perp\cap E_A)+S\big)/S$ be the space of logical classes
that have a representative supported on $A$, and $\ell_S(A)=\dim_{\Fq}\cL_S(A)$. Since
$(S^\perp\cap E_A)\cap S=S\cap E_A$,
\begin{equation}\label{eq:ell}
 \ell_S(A)=\dim_{\Fq}(S^\perp\cap E_A)-\dim_{\Fq}(S\cap E_A).
\end{equation}

\begin{lemma}\label{lem:logical}
Let $S$ be isotropic with $\dim_{\Fq}S=r$, $k=n-r$, and $A\subseteq[n]$.
\begin{enumerate}
\item $\cL_S(\bar A)$ is the orthogonal complement of $\cL_S(A)$ in $\cL(S)$; in particular
$\ell_S(A)+\ell_S(\bar A)=2k$.
\item Let $R$ be a reference system purifying the maximally mixed state $q^{-k}\Pi_S$ of $Q$. Then
$I(A:R)=\ell_S(A)\log q$.
\end{enumerate}
\end{lemma}
\begin{IEEEproof}
Let $s_A=\dim_{\Fq}(S\cap E_A)$. The coordinate projection of $S$ onto $J_A$ has kernel $S\cap E_{\bar A}$,
so its image $S_A'$ has dimension $r-s_{\bar A}$. A vector of $E_A$ is orthogonal to $S$ if and only if
it is orthogonal to $S_A'$, and $E_A$ is nondegenerate by \eqref{eq:split}; so
$\dim_{\Fq}(S^\perp\cap E_A)=2|A|-r+s_{\bar A}$, and by \eqref{eq:ell}
\begin{equation}\label{eq:ell-formula}
 \ell_S(A)=2|A|-r+s_{\bar A}-s_A .
\end{equation}

(1) Adding \eqref{eq:ell-formula} for $A$ and $\bar A$ gives $2n-2r=2k$. Vectors supported on $A$ and on
$\bar A$ are orthogonal, so $\cL_S(\bar A)\subseteq\cL_S(A)^\perp$, and both spaces have dimension
$2k-\ell_S(A)$.

(2) We have $q^{-k}\Pi_S=q^{-n}\sum_{s\in S}\Dh(s)$. The partial trace over $\bar A$ kills every
$\Dh(s)$ with $\supp(s)\not\subseteq A$, so the reduced state on $A$ is
$q^{-|A|}\sum_{s\in S\cap E_A}\Dh(s)|_A$. By Lemma~\ref{lem:lift}, applied on the qudits in $A$, this
is $q^{s_A-|A|}$ times a projection of rank $q^{|A|-s_A}$; its entropy is $(|A|-s_A)\log q$. The
entropy of $R$ is $k\log q$, and that of $AR$ equals that of $\bar A$, because the joint state of
$A$, $\bar A$ and $R$ is pure. Hence
$I(A:R)=\big(|A|-s_A+k-|\bar A|+s_{\bar A}\big)\log q$, which is \eqref{eq:ell-formula}.
\end{IEEEproof}

Part (1) is a form of the cleaning lemma \cite{BravyiTerhal2009}, and part (2) is the stabilizer case
of the information-theoretic description of erasure correction \cite{SchumacherNielsen1996}: by
Lemma~\ref{lem:erasure}, $\bar A$ can recover the code if and only if $\ell_S(A)=0$, if and only if
$\ell_S(\bar A)=2k$.

\subsection{Symplectic matroids}\label{subsec:matroids}

For $X\subseteq J$, define $X^*=\{x^*:x\in X\}$, where $(i^*)^*=i$. A set $X$ is \emph{admissible} if
$X\cap X^*=\emptyset$, and a \emph{transversal} if moreover $|X|=n$. Let $\varphi\colon J\to[n]$ be the map which forgets
the stars. A \emph{symplectic matroid} of rank $r$ on $J$ is a nonempty family $\cB$ of admissible
$r$-sets, the \emph{bases}, satisfying the Gale-maximality axiom of
\cite[Ch.~3]{BorovikGelfandWhite2003}; see also \cite[Def.~2.2]{GangulyPaperI}. Its
\emph{independent} sets are the subsets of bases, its \emph{circuits} are the minimal admissible
sets that are not independent, and its rank function on admissible sets is
$r_{\cS}(X)=\max_{B\in\cB}|X\cap B|$. It is \emph{Lagrangian} if $r=n$; Lagrangian matroids are the
same as Bouchet's delta-matroids \cite{Bouchet1987,BorovikGelfandWhite2003}. The \emph{dual} of a
Lagrangian matroid $L$ has the bases $B^*$, $B\in\cB(L)$, and $L$ is \emph{self-dual} if
$\cB(L)^*=\cB(L)$ \cite{Sarvepalli2011}. The \emph{uniform} symplectic matroid $U^*_{k,n}$ of rank $k$
has as bases all admissible $k$-sets. If $[n]=N_1\sqcup N_2$ and $\cS_i$ is a symplectic matroid on
$J_{N_i}$, the \emph{direct sum} $\cS_1\oplus\cS_2$ is the symplectic matroid on $J$ whose bases are
the sets $B_1\sqcup B_2$ with $B_i\in\cB(\cS_i)$ \cite{BorovikGelfandWhite2003}. If $a\in J$ and $\{a\}$
is independent, the \emph{contraction} $\cS/a$ is the symplectic matroid on $J\setminus\{a,a^*\}$ with
the bases $B\setminus\{a\}$, $a\in B\in\cB(\cS)$ \cite[Lemma~2.8]{GangulyPaperI}.

Let $M\in\Fq^{r\times2n}$ have rank $r$ and an isotropic row space, and for $X\subseteq J$ let
$M[X]$ be the submatrix formed by the columns in $X$. The admissible $r$-sets $X$ with
$\det M[X]\neq0$ are the bases of a symplectic matroid \cite[Ch.~3]{BorovikGelfandWhite2003}, which
depends only on $\row(M)$. We say that $M$, or $\row(M)$, \emph{represents} it. Thus a stabilizer matrix
of an $[[n,n-r]]_q$ code represents a symplectic matroid of rank $r$, and conversely
\cite{Sarvepalli2011}.

\begin{lemma}[{\cite[Lemma~3.1]{GangulyPaperI}}]\label{lem:intrinsic}
If $M$ represents $\cS$, then $\rk M[X]=r_{\cS}(X)$ for every admissible $X$. In particular, the
columns of an admissible set are linearly independent if and only if the set is independent in $\cS$.
\end{lemma}

\subsection{Pair ranks}\label{subsec:pairrank}

\begin{definition}\label{def:pairrank}
The \emph{pair rank} of $M\in\Fq^{r\times2n}$ is the function $\rho_M(A)=\rk M[J_A]$, $A\subseteq[n]$.
\end{definition}

The pair rank depends only on $\row(M)$. It is the rank function of the $2$-polymatroid on $[n]$
formed by the column pairs of $M$, and it does not change when the two columns of each pair are
replaced by an invertible linear combination of them. For $A\neq\emptyset$ the set $J_A$ is not
admissible, so Lemma~\ref{lem:intrinsic} does not express $\rho_M$ in terms of the matroid; whether
the matroid nevertheless determines $\rho_M$ is the subject of Section~\ref{subsec:rigidity}.

\begin{lemma}\label{lem:counting}
Let $S=\row(M)$ be isotropic with $\dim_{\Fq}S=r$, and $A\subseteq[n]$. Then
\begin{gather*} \dim_{\Fq}(S\cap E_A)=r-\rho_M(\bar A),\\ \dim_{\Fq}(S^\perp\cap E_A)=2|A|-\rho_M(A), \end{gather*}
and $\ell_S(A)=2|A|-r-\rho_M(A)+\rho_M(\bar A)$. In particular, if $k\ge1$, then $d(S)$ is the least
$|A|$ with $2|A|-\rho_M(A)>r-\rho_M(\bar A)$. If $S$ is Lagrangian, then
$\rho_M(\bar A)=n-2|A|+\rho_M(A)$ for every $A$.
\end{lemma}
\begin{IEEEproof}
The coordinate projection of $S$ onto $J_A$ has image of dimension $\rho_M(A)$ and kernel
$S\cap E_{\bar A}$; this gives the first formula (with $A$ and $\bar A$ exchanged), and the second
follows as in the proof of Lemma~\ref{lem:logical}. The formula for $\ell_S$ is \eqref{eq:ell}. A
vector of $S^\perp\setminus S$ of weight $d(S)$ is supported on a set of size $d(S)$, which gives the
distance. If $S=S^\perp$, the two dimensions are equal.
\end{IEEEproof}

The dimension and rank identities in Lemma~\ref{lem:counting} hold over any field, by the same
linear-algebraic proof; their coding interpretations use finite fields.

\subsection{Graph states}\label{subsec:graphstates}

A symmetric matrix $A\in\Fq^{n\times n}$ is the adjacency matrix of a weighted graph on $[n]$,
possibly with loops. The row space $W_A$ of $[I_n\mid A]$ is Lagrangian, and its stabilizer state is
the \emph{graph state} of $A$ \cite{SchlingemannWerner2001,HeinEisertBriegel2004}. The
\emph{cut-rank} of $Y\subseteq[n]$ is $\cutrk_A(Y)=\rk A[Y,\bar Y]$ \cite{Oum2005}, where $A[Y,\bar Y]$
is the submatrix with rows in $Y$ and columns in $\bar Y$; by symmetry
$\cutrk_A(Y)=\cutrk_A(\bar Y)$.

\begin{lemma}\label{lem:cutrank}
For $M=[I_n\mid A]$ and $Y\subseteq[n]$, $\rho_M(Y)=|Y|+\cutrk_A(Y)$.
\end{lemma}
\begin{IEEEproof}
The columns indexed by $J_Y$ are the unit vectors $\e_y$ and the columns $A_{\cdot,y}$, $y\in Y$.
Eliminating with the unit vectors leaves the rows $\bar Y$ of the columns $A_{\cdot,Y}$, which have rank
$\cutrk_A(\bar Y)=\cutrk_A(Y)$.
\end{IEEEproof}

\section{Codes from graphical symplectic matroids}\label{sec:graph}

\subsection{Representable graphical symplectic matroids}
Let $G=(V,E)$ be a finite multigraph, loops allowed, with $E=\{e_1,\dots,e_n\}$. The pair $\{i,i^*\}$ is identified with the edge $e_i$; we write $e$ and $e^*$ for the unstarred and the
starred element of the pair of an edge $e$. For an admissible $A\subseteq J$, let $G[A]$ be the spanning subgraph formed by the edges whose pair meets $A$; an edge is \emph{starred} in $A$ if its starred element lies in $A$. Chow \cite{Chow2003} called $A$ independent if every component of $G[A]$ is either a tree or contains exactly one cycle and this cycle has an odd number of starred edges; these sets are the independent sets of a symplectic matroid $\GraphS{G}$ \cite[Thm.~2]{Chow2003}, \cite[Cor.~3.5]{GangulyPaperI}. A bridge is a non-loop edge lying on no cycle; $G^\circ$ denotes $G$ with all loops and bridges deleted, and $\nuo(G)$ is the maximum number
of vertex-disjoint cycles. We use the following results of \cite{GangulyPaperI}.

\begin{theorem}[{\cite{GangulyPaperI}}]\label{thm:paperI}
Let $G$ be a graph and $\F$ a field.
\begin{enumerate}
\item $r(\GraphS{G})=|V|-\tau(G)$, where $\tau(G)$ is the number of acyclic components of $G$ \cite[Lemma~2.12]{GangulyPaperI}.
\item \emph{(Normal form \cite[Thm.~3.8]{GangulyPaperI}.)} Let $G$ be connected and contain a cycle, fix an orientation and a root vertex, and let $D\in\F^{(|V|-1)\times n}$ be the reduced
incidence matrix, with columns $\bd_e$. Every representation of $\GraphS{G}$ over $\F$ is transformed by left multiplication by an invertible matrix and by pair scalings
$(x_e,x_{e^*})\mapsto(c_e^{-1}x_e,c_ex_{e^*})$ with $c_e\in\F^\times$, into a representation $H=\left[\begin{smallmatrix}D&Z_1\\0&z\end{smallmatrix}\right]$ in which $z$ is an $\F$-flow on $G$ that is nonzero exactly on the edges lying on cycles, and $DZ_1^T$ is symmetric. For $z_e\neq0$ put $w_e=z_e^{-1}Z_{1,e}$, so that the columns of $e$ and $e^*$ are $(\bd_e;0)$ and $z_e(w_e;1)$.
\item \emph{(Two-star relations \cite[Lemma~3.9]{GangulyPaperI}.)} In such a representation, if
$C$ is a cycle of non-loop edges and $e\neq f\in C$, then
$w_e-w_f\in\Span_{\F}\{\bd_g:g\in C\setminus\{e,f\}\}$.
\item If $\operatorname{Char}\F\neq2$, then $\GraphS{G}$ is $\F$-representable if and only if every edge that lies on a cycle and is not a loop has a parallel mate \cite[Thm.~5.1]{GangulyPaperI}.
\item If $\operatorname{Char}\F=2$, $G$ is loopless and $\nuo(G)\le1$, then $\GraphS{G}$ is $\F$-representable if and only if $G^\circ$ has a nowhere-zero $(\F,+)$-flow; if $G$ is connected and contains a cycle, and $\lambda$ is an $\F$-flow on $G$ that is nonzero exactly on the edges that are not
bridges, then the \emph{one-row lift} $H_\lambda=\left[\begin{smallmatrix}D&D\,\mathrm{diag}(c)\\0&\lambda\end{smallmatrix}\right]$, with
$c_e=\lambda_e$ for non-bridges and $c_e=1$ for bridges, represents $\GraphS{G}$ \cite[Prop.~6.1, Thm.~6.2]{GangulyPaperI}.
\item If $G$ is simple and $3$-connected, then $\GraphS{G}$ is $\F$-representable if and only if
$\operatorname{Char}\F=2$, $\nuo(G)\le1$, and either $|\F|>2$ or $G\cong K_5$
\cite[Thm.~6.11]{GangulyPaperI}. The $3$-connected simple graphs with $\nuo(G)\le1$ are $K_5$,
$K_5$ minus an edge, the wheels, and the graphs obtained from $K_{3,p}$, $p\ge3$, by adding edges
inside the part of size three \cite{Lovasz1965,Dirac1963}.
\end{enumerate}
\end{theorem}

In (2) we identify $\F^{V\setminus\{\text{root}\}}$ with the space of vectors of $\F^V$ with
coordinate sum zero; then $\bd_e=\one_s-\one_t$ for an edge from $s$ to $t$, where $\one_s$ is the unit
vector of $s$. For $U\subseteq V$ put
\[ \Fz{U}=\Big\{x\in\F^V:\ x_v=0\text{ for }v\notin U,\ \textstyle\sum_{v\in U}x_v=0\Big\}. \]
If $K$ is a connected subgraph, then $\Span_{\F}\{\bd_g:g\in E(K)\}=\Fz{V(K)}$, because a spanning
tree of $K$ gives $|V(K)|-1$ linearly independent vectors in this space of dimension $|V(K)|-1$.

\subsection{Parameters and a lower bound}

\begin{proposition}\label{prop:graph-params}
If $H\in\Fq^{r\times2n}$ represents $\GraphS{G}$, its code has parameters
$[[\,|E|,\,|E|-|V|+\tau(G)\,]]_q$. If $G$ is connected and contains a cycle, these are
$[[\,|E|,\,|E|-|V|\,]]_q$. A distance as in \eqref{eq:distance} is assigned when the number of
logical qudits is positive.
\end{proposition}
\begin{IEEEproof}
By Theorem~\ref{thm:paperI}(1), $r=|V|-\tau(G)$, and the code has $n-r$ logical qudits.
\end{IEEEproof}

\begin{theorem}\label{thm:weight-one}
Let $H$ represent $\GraphS{G}$, let $S=\mathrm{row}(H)$, and let $e$ be an edge that is neither a
loop nor a bridge. Then $\rho_H(\{e\})=2$, so no nonzero vector of $S^\perp$ is supported on the
qudit $e$. In particular,
if $G$ is loopless and bridgeless and the code has at least one logical qudit, then $d\ge2$.
\end{theorem}
\begin{IEEEproof}
Let $C$ be a cycle through $e$; it has length at least two. The set $(C\setminus\{e\})\cup\{e^*\}$
spans a unicyclic subgraph whose cycle has one starred edge, so it is independent, while
$C=(C\setminus\{e\})\cup\{e\}$ is a cycle without starred edges, hence dependent. By
Lemma~\ref{lem:intrinsic}, the column of $e$ lies in the $\Fq$-span of the columns of $C\setminus\{e\}$
and the column of $e^*$ does not. Both columns are nonzero, since $\{e\}$ and $\{e^*\}$ are
independent. Hence they are linearly independent and $\rho_H(\{e\})=2$. Lemma~\ref{lem:counting}
gives $\dim_{\Fq}(S^\perp\cap E_{\{e\}})=0$.
\end{IEEEproof}

\subsection{The distance bound}

\begin{theorem}\label{thm:barrier}
Let $G$ be a loopless multigraph, $\F$ any field, and $H$ any representation of $\GraphS{G}$ over
$\F$. Suppose that $G$ has distinct edges $e,f$ such that $G-\{e,f\}$ is connected and contains a
cycle, and such that
\begin{enumerate}
\item[(a)] $e$ and $f$ are parallel; or
\item[(b)] $e=uw$ and $f=wx$ with $u\neq x$, the vertices $u$ and $x$ are connected in $G-w$, and no
vertex other than $u,x$ separates $u$ from $x$ in $G-w$.
\end{enumerate}
Then $n-\dim_{\F}S\ge2$, where $S=\row_{\F}(H)$, and $S^\perp/S$ has a nonzero class supported
on $\{e,f\}$. In particular, when $\F=\Fq$, the code has at least two logical qudits and distance at
most two.
\end{theorem}
\begin{IEEEproof}
Let $S=\mathrm{row}(H)$. $G$ is connected and contains a cycle, so $r=|V|$; and $G-\{e,f\}$ is
connected and contains a
cycle, so it has at least $|V|$ edges and $n\ge|V|+2$. Fix an orientation in which, in case (b),
$e$ is directed from $u$ to $w$ and $f$ from $w$ to $x$. Row operations and pair scalings preserve
supported quotient dimensions and pair ranks (and hence code parameters over finite fields), so by Theorem~\ref{thm:paperI}(2) we may take $H$ in normal
form. Neither $e$ nor $f$ is a bridge, so $z_e,z_f\neq0$.

We claim that $w_e-w_f\in\Span_{\F}\{\bd_e,\bd_f\}$. In case (a), $\{e,f\}$ is a cycle of length two
and Theorem~\ref{thm:paperI}(3) gives $w_e=w_f$. In case (b), for every $u$--$x$ path $P$ in
$G-w$, $P\cup\{e,f\}$ is a cycle, and
Theorem~\ref{thm:paperI}(3) gives $w_e-w_f\in\Span_{\F}\{\bd_g:g\in P\}=\Fz{V(P)}$. By hypothesis every
vertex other than $u,x$ is avoided by some such path, so
$w_e-w_f\in\bigcap_P\Fz{V(P)}=\Fz{\{u,x\}}=\Span_{\F}\{\one_u-\one_x\}=\Span_{\F}\{\bd_e+\bd_f\}$.

The columns indexed by $J_{\{e,f\}}$ are $(\bd_e;0)$, $(\bd_f;0)$, $z_e(w_e;1)$ and $z_f(w_f;1)$.
By the claim, $z_e^{-1}\mathrm{col}(e^*)-z_f^{-1}\mathrm{col}(f^*)=(w_e-w_f;0)$ lies in the $\F$-span of
the first two, so $\rho_H(\{e,f\})\le3$. On the other hand, $G-\{e,f\}$ contains a connected
spanning subgraph with exactly one cycle; starring one edge of that cycle gives a basis of
$\GraphS{G}$ that avoids the pairs of $e$ and $f$, so $\rho_H(E\setminus\{e,f\})=r$. By
Lemma~\ref{lem:counting}, $\dim_{\F}(S^\perp\cap E_{\{e,f\}})\ge1>0=\dim_{\F}(S\cap E_{\{e,f\}})$.
\end{IEEEproof}

\begin{corollary}\label{cor:3conn}
Let $G$ be a $3$-connected simple graph with $|V|\ge4$. For every field $\Fq$ over which
$\GraphS{G}$ is representable and every representation, the code is an
$[[\,|E|,\,|E|-|V|,\,2\,]]_q$ code.
\end{corollary}
\begin{IEEEproof}
Let $w$ be a vertex and $e=wu$, $f=wx$ two edges at $w$. Since $G-w$ is $2$-connected, condition (b)
of Theorem~\ref{thm:barrier} holds. Since $G$ is $3$-edge-connected, $G-\{e,f\}$ is connected; it
has $|E|-2\ge\frac32|V|-2\ge|V|$ edges, so it contains a cycle. Thus $d\le2$, and $d\ge2$ by
Theorem~\ref{thm:weight-one}.
\end{IEEEproof}

\begin{corollary}\label{cor:oddchar-codes}
Let $\operatorname{Char}\Fq\neq2$, let $G$ be loopless and bridgeless, and suppose that
$\GraphS{G}$ is $\Fq$-representable. If some pair of parallel edges $e,f$ leaves $G-\{e,f\}$
connected and not a tree, then every representation gives $d=2$.
\end{corollary}
\begin{IEEEproof}
Apply Theorem~\ref{thm:barrier}(a) and Theorem~\ref{thm:weight-one}. By
Theorem~\ref{thm:paperI}(4), every edge of $G$ has a parallel mate, so the hypothesis concerns only
the global structure of $G$.
\end{IEEEproof}

\begin{example}[The girth does not bound the distance]\label{ex:K4}
In \cite[Sec.~III]{Sarvepalli2011} the code of a representable $\GraphS{G}$ is stated to have distance
at least the length of a shortest cycle of $G$. Let $G=K_4$ on $\{0,1,2,3\}$, with the edges ordered
$01,02,03,12,13,23$, and let $\F_4=\F_2(\vartheta)$ with $\vartheta^2=\vartheta+1$ and
$\bar\vartheta=\vartheta^2$. Give the perfect matchings $\{01,23\}$, $\{02,13\}$, $\{03,12\}$ the values
$1$, $\vartheta$, $\bar\vartheta$; since $1+\vartheta+\bar\vartheta=0$, this is a nowhere-zero
$\F_4$-flow $\lambda$. With the root $0$, its one-row lift (Theorem~\ref{thm:paperI}(5)) is
{\small\[
H_\lambda=\left[\begin{array}{@{}cccccc|cccccc@{}}
1&0&0&1&1&0&1&0&0&\bar\vartheta&\vartheta&0\\
0&1&0&1&0&1&0&\vartheta&0&\bar\vartheta&0&1\\
0&0&1&0&1&1&0&0&\bar\vartheta&0&\vartheta&1\\
0&0&0&0&0&0&1&\vartheta&\bar\vartheta&\bar\vartheta&\vartheta&1
\end{array}\right],
\]}
and of the $\binom64 2^4=240$ admissible $4$-sets exactly the $120$ independent ones have nonzero
minors. Its code is $[[6,2,2]]_4$: the vector with $X$-part $(1,\vartheta)$ and $Z$-part
$(\vartheta,\vartheta)$ on the edges $13,23$, and zero elsewhere, lies in $S^\perp\setminus S$. In fact
every pair of edges carries a nontrivial logical operator. The girth of $K_4$ is $3$. By
Corollary~\ref{cor:3conn}, every representation of $\GraphS{K_4}$ over a finite field has distance two.
\end{example}

\begin{example}\label{ex:theta}
Let $\Theta=K_{2,3}$ have poles $0,1$, inner vertices $2,3,4$, and edges ordered
$02$, $21$, $03$, $31$, $04$, $41$. Over $\F_4$, with $\bar\vartheta=\vartheta^2$, the matrix
{\small\[
H_\Theta=\left[\begin{array}{@{}c@{\,}c@{\,}c@{\,}c@{\,}c@{\,}c|c@{\,}c@{\,}c@{\,}c@{\,}c@{\,}c@{}}
0&1&0&1&0&1&\vartheta&1&0&\vartheta&\vartheta&\bar\vartheta\\
1&1&0&0&0&0&\bar\vartheta&\bar\vartheta&0&0&0&0\\
0&0&1&1&0&0&\bar\vartheta&\bar\vartheta&\bar\vartheta&\bar\vartheta&\vartheta&\vartheta\\
0&0&0&0&1&1&0&0&0&0&1&1\\
0&0&0&0&0&0&1&1&\vartheta&\vartheta&\bar\vartheta&\bar\vartheta
\end{array}\right]
\]}
is isotropic and represents $\GraphS{\Theta}$: of the $\binom65 2^5=192$ admissible $5$-sets,
exactly the $96$ independent ones have nonzero minors. Its code is $[[6,1,3]]_4$, with weight
enumerators
\begin{gather*} W_S(z)=1+9z^2+27z^4+288z^5+699z^6,\\ W_{S^\perp}(z)=1+9z^2+168z^3+963z^4+3672z^5+11571z^6 \end{gather*}
(see Section~\ref{sec:enum}). Hypothesis (b)
of Theorem~\ref{thm:barrier} holds only at the inner vertices, where $G-\{e,f\}$ is disconnected.
The claim here concerns graphical realization of these parameters, not novelty of the code
up to equivalence. The example shows that the distance bound is not universal.
\end{example}

\subsection{Matroid operations and associated quantum codes}

Sarvepalli \cite[Sec.~III-B]{Sarvepalli2011} relates contraction, truncation and the Higgs lift of
symplectic matroids \cite{BorovikGelfandWhite2003} to operations on codes. We make this precise,
using the contraction $\cS/a$ defined in Section~\ref{subsec:matroids}.

\begin{proposition}[Contraction]\label{prop:contraction}
Let $H\in\Fq^{r\times2n}$ represent $\cS$, with code $[[n,k,d]]_q$, $k=n-r\ge1$, and let $a\in J$
have a nonzero column. Row-reduce $H$ so that this column is the first unit vector, and let $H/a$ be
obtained by deleting the first row and the two columns of the pair of $a$. Then $H/a$ is isotropic of
rank $r-1$, it represents $\cS/a$, and its code is an $[[n-1,k,d']]_q$ code with $d'\ge d-1$.
\end{proposition}
\begin{IEEEproof}
The first three statements are \cite[Lemma~3.3]{GangulyPaperI}. Let $a$ belong to the pair of qudit
$i$, and let $v'\in(S/a)^\perp\setminus(S/a)$, where $S/a=\mathrm{row}(H/a)$. Extend $v'$ to $v\in\Fq^{2n}$ by setting the coordinate $a$ equal to $0$ and choosing the coordinate $a^*$ so that $\omega(v,h_1)=0$, where $h_1$ is the first row of $H$; this is possible because $h_1$ has the entry
$1$ at $a$. The other rows are zero at $a$, so $v$ is also orthogonal to them, and $v\in S^\perp$. If $v=\sum_jc_jh_j$, the coordinate $a$ gives $c_1=0$, and then $v'\in S/a$, a contradiction. Hence $v\in S^\perp\setminus S$ and $d\le\wt(v)\le\wt(v')+1$.
\end{IEEEproof}

\begin{proposition}[Truncation]\label{prop:truncation}
Let $H\in\Fq^{r\times2n}$ represent $\cS$, let $1\le t<r$, and let $\cS^{(t)}$ be the truncation
to rank $t$, whose bases are the independent $t$-sets of $\cS$ \cite[Lemma~2.8]{GangulyPaperI}. Let
$\F=\mathbb F_{q^\ell}$ be a finite extension of $\Fq$.
\begin{enumerate}
\item For every $R\in\F^{t\times r}$ of rank $t$, the matrix $RH$ is isotropic and
$\row_{\F}(RH)\subseteq\row_{\F}(H)$, where $\row_{\F}$ denotes the $\F$-span of the rows; hence, with compatible stabilizer lifts, the $[[n,n-t]]_{|\F|}$ code of $RH$ contains
the code of $H$ regarded over $\F$. If $|\F|$ is large enough, $R$ can be chosen so that $RH$
represents $\cS^{(t)}$.
\item In general $\F\neq\Fq$ is necessary. The matrix $[I_5\mid I_5]$ represents $U^*_{5,5}$ over
$\F_2$, whose truncation $U^*_{3,5}$ to rank $3$, that is, all admissible $3$-sets are bases, is not $\F_2$-representable, while $[V\mid V]$ represents $U^*_{3,5}$ over $\F_4$ when the columns of $V\in\F_4^{3\times5}$ form an arc of $\mathrm{PG}(2,4)$.
\end{enumerate}
\end{proposition}
\begin{IEEEproof}
(1) A subspace of an isotropic subspace is isotropic. Restricting a lift of $\row_{\F}(H)$ to
$\row_{\F}(RH)$ gives the claimed codespace inclusion. For an admissible $t$-set $A$, the
Cauchy--Binet formula gives $\det(RH)[A]=\sum_T\det R[\cdot,T]\,\det H[T,A]$, summed over the
$t$-subsets $T$ of the rows. The minors $\det R[\cdot,T]$ are linearly independent polynomials in
the entries of $R$, so $\det(RH)[A]$ is a nonzero polynomial if and only if $\rk H[A]=t$, that is,
by Lemma~\ref{lem:intrinsic}, if and only if $A$ is independent in $\cS$. If $|\F|$ exceeds the
degree of the product of these finitely many nonzero polynomials, some $R$ avoids all their zeros;
then $RH$ has rank $t$, and its bases are exactly the independent $t$-sets of $\cS$. (2) If $H'$
represented
$U^*_{3,5}$ over $\F_2$, the five columns $x_1,\dots,x_5$ indexed by $[5]$ would have every three independent.
Normalize three to the unit vectors of $\F_2^3$. Each remaining column must have all three
coordinates nonzero, since it must be independent of every pair of unit vectors. Thus both
remaining columns would equal $(1,1,1)^T$, a contradiction. Over $\F_4$ one may take
\[ V=\begin{bmatrix}1&0&0&1&1\\0&1&0&1&\vartheta\\0&0&1&1&\vartheta^2\end{bmatrix},
\qquad \vartheta^2=\vartheta+1. \]
Its ten $3\times3$ minors are nonzero. For $[V\mid V]$, isotropy is $VV^T=VV^T$, and every
admissible $3$-set selects three distinct columns of $V$.
\end{IEEEproof}

\begin{remark}[Stabilizer extension and Higgs lifts]\label{rem:higgs}
For $k\ge2$ and $v\in S^\perp\setminus S$, the space $S'=S+\Fq v$ is isotropic of dimension $r+1$.
With compatible lifts its code is a subcode of $Q(S)$ and has parameters $[[n,k-1,d']]_q$,
$d'\ge d$, because $S'^\perp\setminus S'\subseteq S^\perp\setminus S$.
Sarvepalli \cite[Sec.~III-B]{Sarvepalli2011} associates this parameter change with a Higgs lift.
The choice of $v$ is additional data and does not specify a canonical operation on the abstract
symplectic matroid. Ordinary Higgs lifts and truncated elongations are treated separately in
Appendix~\ref{app:envelopes}; their isotropic representability requires verification.
\end{remark}

\begin{remark}[Answer to Question~2]\label{rem:Q2}
Let $G$ be a simple connected graph with a cycle. Then $\GraphS{G}$ is representable only in
characteristic $2$ (Theorem~\ref{thm:paperI}(4)), and simple unicyclic graphs give $[[n,0]]_q$ codes.
If $G$ is $3$-connected, Theorem~\ref{thm:paperI}(6) lists the graphs that give codes at all: $K_5$,
over every finite field of characteristic $2$, and $K_5$ minus an edge, the wheels and the graphs
built from $K_{3,p}$, over finite fields of characteristic $2$ with more than two elements. By
Corollary~\ref{cor:3conn}, every representation of each of them gives distance exactly $2$; for
instance the wheel with $k$ spokes gives $[[2k,k-1,2]]_q$ codes and $K_5$ gives $[[10,5,2]]_q$ codes.
These codes detect arbitrary errors of weight one but cannot correct all errors of weight one. Every other $3$-connected simple graph, such as
the Petersen graph \cite[Cor.~6.12]{GangulyPaperI}, gives no code over any field, because its
matroid is not representable. Larger distance
therefore requires graphs of connectivity at most $2$ in which the configurations of
Theorem~\ref{thm:barrier} are absent, such as subdivisions (Example~\ref{ex:theta}). Whether the
distance is unbounded within such families is open (Section~\ref{sec:concl}).
\end{remark}

\section{Quantum secret-sharing from represented Lagrangian matroids}\label{sec:qss}

\subsection{Quantum secret-sharing Schemes and access structures}

\begin{definition}\label{def:qss}
Let $P$ be a finite player set and let $\mathcal H_i$ be a finite-dimensional share space for each
$i\in P$. A quantum secret-sharing encoding of a secret in $\C^q$ is a completely positive
trace-preserving map $\mathcal E:\mathcal B(\C^q)\to\mathcal B(\bigotimes_{i\in P}\mathcal H_i)$.
It is pure if $\mathcal E(\varrho)=V\varrho V^\dagger$ for an isometry $V$.
For $A\subseteq P$, put $\mathcal E_A=\Tr_{P\setminus A}\circ\mathcal E$.
The set $A$ is authorized if a channel $\mathcal R_A$ satisfies
$\mathcal R_A\circ\mathcal E_A=\mathrm{id}_{\mathcal B(\C^q)}$; it is forbidden if
$\mathcal E_A(\varrho)$ is independent of $\varrho$; and it is intermediate otherwise.
The scheme is perfect if no set is intermediate. For pure schemes we write
$\varrho_A=\mathcal E_A(\varrho)$. Our dealer-shortened schemes have $\mathcal H_i=\C^q$;
hybrid schemes also include a classical register in each share.
\end{definition}

The authorized sets form a monotone family, and no two of them are disjoint, by the no-cloning theorem. A nonempty family $\Gamma$ of subsets of $P$ is called a \emph{quantum access structure} if it is monotone, that is, if $A\in\Gamma$ and $A\subseteq A'\subseteq P$ then $A'\in\Gamma$, and any two members of $\Gamma$, not necessarily distinct, have nonempty intersection. Every quantum access structure is the family of authorized sets of some perfect scheme with suitable share dimensions, possibly with mixed-state encodings, and for a pure-state scheme a set is authorized if and only if its complement is forbidden \cite{CleveGottesmanLo1999,Gottesman2000,Smith2000}. If $V$ encodes into a stabilizer code with stabilizer $S$ and $k=1$, Lemmas~\ref{lem:erasure} and~\ref{lem:logical} show that $A$ is authorized if and only if $\ell_S(A)=2$ and forbidden if and only if $\ell_S(A)=0$.

\subsection{Circuit-induced access structures}

Let $L$ be a Lagrangian matroid on $J$, and let $\cC(L)$ be its set of circuits; following
\cite{Sarvepalli2011}, the sets $C^*$, $C\in\cC(L)$, are called the \emph{cocircuits} of $L$. Fix a
\emph{dealer} $d\in[n]$, let $P_d=[n]\setminus\{d\}$ be the set of players and $J_d=\{d,d^*\}$, and put
\[ \cC_d(L)=\{C\in\cC(L):\ C\cap J_d\neq\emptyset\}. \]
A circuit $C\in\cC_d(L)$ is admissible, so it contains exactly one element of $J_d$, which we denote by
$d_C$. Its \emph{player support} is
\[ B_d(C)=\varphi(C\setminus\{d_C\})=\varphi(C)\setminus\{d\}\subseteq P_d, \]
and we write $\mathcal F_{L,d}=\{B_d(C):C\in\cC_d(L)\}$.

\begin{definition}[\cite{SarvepalliRaussendorf2010,Sarvepalli2011}]\label{def:circ}
Let $\Gmin_{L,d}$ be the set of inclusion-minimal members of $\mathcal F_{L,d}$. The
\emph{circuit-induced access structure} is
\[ \Gcirc_{L,d}=\{A\subseteq P_d:\ B\subseteq A\text{ for some }B\in\Gmin_{L,d}\}. \]
The matroid $L$ is \emph{secret-sharing} if $\Gcirc_{L,d}$ is a quantum access structure for every
$d\in[n]$.
\end{definition}

The following elementary lemma reduces the no-cloning property to the generating sets.

\begin{lemma}[Monotone closures]\label{lem:closure}
Let $\mathcal F$ be a finite family of subsets of a finite set $P$, let $\Gamma_0$ be the set of its
inclusion-minimal members, and let $\Gamma=\{A\subseteq P:B\subseteq A\text{ for some }B\in\Gamma_0\}$.
Then $\Gamma_0$ is an antichain, every member of $\mathcal F$ contains a member of $\Gamma_0$, and $\Gamma$
is the smallest monotone family containing $\mathcal F$. Moreover, the following are equivalent:
\textup{(a)} $\Gamma$ is a quantum access structure; \textup{(b)} $\Gamma_0\neq\emptyset$ and
$B\cap B'\neq\emptyset$ for all $B,B'\in\Gamma_0$; \textup{(c)} $\mathcal F\neq\emptyset$ and
$F\cap F'\neq\emptyset$ for all $F,F'\in\mathcal F$.
\end{lemma}
\begin{IEEEproof}
Since $\mathcal F$ is finite, every $F\in\mathcal F$ contains an inclusion-minimal member of $\mathcal F$,
and two distinct minimal members are incomparable. Hence $\mathcal F\subseteq\Gamma$, and every monotone
family containing $\mathcal F$ contains $\Gamma$. (a)$\Rightarrow$(c) because $\mathcal F\subseteq\Gamma$.
(c)$\Rightarrow$(b) because $\emptyset\neq\Gamma_0\subseteq\mathcal F$. (b)$\Rightarrow$(a): $\Gamma$ is
monotone and nonempty, and if $A,A'\in\Gamma$ contain $B,B'\in\Gamma_0$, then
$A\cap A'\supseteq B\cap B'\neq\emptyset$.
\end{IEEEproof}

\begin{proposition}\label{prop:nocloning}
Let $L$ be a Lagrangian matroid and $d\in[n]$. Then $\Gcirc_{L,d}$ is a quantum access structure if and
only if $\cC_d(L)\neq\emptyset$ and
\[ \big(\varphi(C)\cap\varphi(C')\big)\setminus\{d\}\neq\emptyset \]
for all $C,C'\in\cC_d(L)$, including $C=C'$.
\end{proposition}
\begin{IEEEproof}
For $C,C'\in\cC_d(L)$ we have $B_d(C)\cap B_d(C')=\big(\varphi(C)\cap\varphi(C')\big)\setminus\{d\}$. Apply
Lemma~\ref{lem:closure} with $\mathcal F=\mathcal F_{L,d}$, $\Gamma_0=\Gmin_{L,d}$ and $\Gamma=\Gcirc_{L,d}$;
condition (c) is the stated condition. For $C=C'$ it says $B_d(C)\neq\emptyset$, that is, $C$ is not
one of the one-element circuits $\{d\}$, $\{d^*\}$.
\end{IEEEproof}

\begin{corollary}\label{cor:secret-sharing}
A Lagrangian matroid $L$ is secret-sharing if and only if, for every $d\in[n]$, $\cC_d(L)\neq\emptyset$
and $\big(\varphi(C)\cap\varphi(C')\big)\setminus\{d\}\neq\emptyset$ for all $C,C'\in\cC_d(L)$.
\end{corollary}
\begin{IEEEproof}
Apply Proposition~\ref{prop:nocloning} to each dealer $d\in[n]$.
\end{IEEEproof}

For representable matroids, circuits correspond to vectors of the representing subspace. Here and below, $W=\row(M)$ is a Lagrangian subspace of $\Fq^{2n}$ with generator matrix $M\in\Fq^{n\times2n}$, and $\ker M=\{c\in\Fq^{2n}:Mc^T=0\}$ is the space of linear relations among the columns of $M$; thus $c\in\ker M$ means $\sum_{j\in J}c_j\,M[\{j\}]=0$.

\begin{lemma}\label{lem:circuit-vector}
The map $T:W\to\ker M$, $T(w)=w\Omega_n$, is an $\Fq$-linear isomorphism, and $\supp_J(T(w))=\supp_J(w)^*$. Consequently, if $W$ represents $L$, then for every circuit $C$ of $L$ there is $w_C\in W$ with $\supp_J(w_C)=C^*$.
\end{lemma}
\begin{IEEEproof}
For $w\in W$, isotropy gives $M\Omega_nw^T=0$, and $\Omega_n^T=-\Omega_n$, so
$M(w\Omega_n)^T=-M\Omega_nw^T=0$. The map is injective because $\Omega_n$ is invertible, and both spaces
have dimension $n$ over $\Fq$, since $\rk M=n$. For $w=(x\mid z)$ we have $w\Omega_n=(-z\mid x)$, which
gives the statement on supports. If $C$ is a circuit, its columns are linearly dependent while those of
every proper subset are independent, by Lemma~\ref{lem:intrinsic}; so some relation $c\in\ker M$ has
$\supp_J(c)=C$. Put $w_C=T^{-1}(c)$.
\end{IEEEproof}

For ordinary matroids, a circuit and a cocircuit never meet in exactly one element
\cite[Prop.~2.1.11]{Oxley2011}. Sarvepalli uses this fact for Lagrangian matroids in
\cite[Thm.~7]{Sarvepalli2011}. For represented Lagrangian matroids it follows from isotropy.

\begin{lemma}\label{lem:orthogonality}
Let $W$ represent $L$, and let $C,D\in\cC(L)$. Then $|C\cap D^*|\neq1$.
\end{lemma}
\begin{IEEEproof}
Take $w=w_C$ and $w'=w_D$ as in Lemma~\ref{lem:circuit-vector}, so $\supp_J(w)=C^*$ and
$\supp_J(w')=D^*$, and write
$0=\omega(w,w')=\sum_{i\in[n]}\big(w_iw'_{i^*}-w_{i^*}w'_i\big)$. Here $w_iw'_{i^*}\neq0$ exactly when
$i^*\in C$ and $i\in D$, that is, $i^*\in C\cap D^*$; and $w_{i^*}w'_i\neq0$ exactly when $i\in C\cap D^*$.
As $C$ is admissible, at most one of these products is nonzero for each $i$, so the term of $i$ is
nonzero exactly when $J_{\{i\}}\cap C\cap D^*\neq\emptyset$. If $|C\cap D^*|=1$, exactly one term is
nonzero, and $\omega(w,w')\neq0$, a contradiction.
\end{IEEEproof}

\begin{theorem}\label{thm:selfdual-ss}
Let $L$ be a self-dual Lagrangian matroid represented by a Lagrangian $W\le\Fq^{2n}$, and $d\in[n]$. Then
$\big(\varphi(C)\cap\varphi(C')\big)\setminus\{d\}\neq\emptyset$ for all $C,C'\in\cC_d(L)$. Hence
$\Gcirc_{L,d}$ is a quantum access structure whenever $\cC_d(L)\neq\emptyset$, and $L$ is secret-sharing
whenever every dealer lies on a circuit.
\end{theorem}
\begin{IEEEproof}
Since ${}^*$ maps $\cB(L)$ onto itself, it maps independent sets to independent sets and hence circuits
to circuits. For $C'\in\cC_d(L)$ we have $C'^*\in\cC_d(L)$ and $\varphi(C'^*)=\varphi(C')$, so we may
replace $C'$ by $C'^*$ and assume $d_C=d_{C'}=a$. Suppose that
$\big(\varphi(C)\cap\varphi(C')\big)\setminus\{d\}=\emptyset$. If $x\in C\cap C'$ and $x\neq a$, then
$x\notin J_d$ and $\varphi(x)$ lies in this set; hence $C\cap C'=\{a\}$. Now $E=C'^*$ is a circuit and
$C'=E^*$, so $|C\cap E^*|=1$, contradicting Lemma~\ref{lem:orthogonality}. The rest follows from
Proposition~\ref{prop:nocloning} and Corollary~\ref{cor:secret-sharing}.
\end{IEEEproof}

Theorem~\ref{thm:selfdual-ss} makes explicit the circuit-nonemptiness hypothesis required when
applying \cite[Thm.~7]{Sarvepalli2011} with the nonempty access structures of Definition~\ref{def:circ}. Over $\F_2$,
a dealer of a self-dual $L$ lies on a circuit exactly when it is surjective, by
Theorem~\ref{thm:selfdual} and Proposition~\ref{prop:nonvacuity} below.

A vector of $W$ may have a support containing both $j$ and $j^*$; such a support is not admissible
and is not related to a circuit. The next proposition says that this is the only obstruction. For
$w\in\Fq^{2n}$ we write $w=(a,p)\in E_d\oplus E_{P_d}$, with $a\in E_d=E_{\{d\}}$ and $p\in E_{P_d}$,
and $\pi_d(w)=a$, the \emph{dealer part} of $w$.

\begin{proposition}\label{prop:admissible-lift}
Let $W$ represent $L$, and $A\subseteq P_d$. Then $A\in\Gcirc_{L,d}$ if and only if there is
$w\in W\cap(E_d\oplus E_A)$ with $\pi_d(w)\neq0$ and $\supp_J(w)$ admissible.
\end{proposition}
\begin{IEEEproof}
If $B_d(C)\subseteq A$ with $C\in\cC_d(L)$, the vector $w_C$ of Lemma~\ref{lem:circuit-vector} has
admissible support $C^*\subseteq J_{A\cup\{d\}}$, and $d_C^*\in C^*$, so $\pi_d(w_C)\neq0$.
Conversely, let $w$ be as stated and $X=\supp_J(w)^*$. Then $X$ is admissible, it is the support of the
relation $T(w)$, and it contains exactly one element $x$ of $J_d$. Since the coefficient of $x$ in
$T(w)$ is nonzero, the column of $x$ lies in the $\Fq$-span of the columns of $X\setminus\{x\}$. Choose
$Y\subseteq X\setminus\{x\}$ whose columns form a basis of this span, and let $Y'\subseteq Y$ be the set of
columns with nonzero coefficient in the (unique) expression of the column of $x$ in this basis. Then the
columns of $\{x\}\cup Y'$ are linearly dependent, the columns of $Y'$ are independent, and for
$y\in Y'$ the columns of $\{x\}\cup Y'\setminus\{y\}$ are independent, since a relation among them
would involve $x$ and give a second expression of the column of $x$. By Lemma~\ref{lem:intrinsic}, the
admissible set $C=\{x\}\cup Y'$ is therefore a circuit of $L$. It meets $J_d$, and
$B_d(C)\subseteq\varphi(X)\setminus\{d\}\subseteq A$.
\end{IEEEproof}

For binary graph states, the circuits, and hence $\Gcirc$, can be read off the graph. Let $\Gamma$ be a
simple graph on $[n]$ with adjacency matrix $A$ over $\F_2$, and let $N(v)$ be the set of neighbours of
$v$. For $Y\subseteq[n]$, the \emph{odd neighbourhood} of $Y$ is $\mathcal{O}(Y)=\{v\in[n]:|N(v)\cap Y|\text{ is odd}\}$, and its \emph{support} is $\supp_\Gamma(Y)=Y\cup \mathcal{O}(Y)$. A nonempty $Y$ is \emph{Eulerian} if $Y\cap \mathcal{O}(Y)=\emptyset$, that is, if every vertex of $Y$ has even degree in the induced subgraph $\Gamma[Y]$, and we call an Eulerian $Y$ is \emph{primitive} if no nonempty $Z\subsetneq Y$ satisfies $\mathcal{O}(Z)\subseteq \mathcal{O}(Y)$. Let $\mathcal P(\Gamma)$ be the set of primitive Eulerian sets, and for $Y\subseteq[n]$ put $C_\Gamma(Y)=\mathcal{O}(Y)\sqcup Y^*\subseteq J$; this set is admissible if and only if $Y\cap \mathcal{O}(Y)=\emptyset$.

\begin{theorem}\label{thm:graph-circuits}
Let $L(\Gamma)$ be the Lagrangian matroid of $[I_n\mid A]$ over $\F_2$, and let $d\in[n]$.
\begin{enumerate}
\item $\cC(L(\Gamma))=\{C_\Gamma(Y):Y\in\mathcal P(\Gamma)\}$.
\item For $Y\in\mathcal P(\Gamma)$, $d^*\in C_\Gamma(Y)$ if and only if $d\in Y$, and $d\in C_\Gamma(Y)$ if
and only if $d\in \mathcal{O}(Y)$; so $C_\Gamma(Y)\in\cC_d(L(\Gamma))$ if and only if $d\in\supp_\Gamma(Y)$, and then $B_d(C_\Gamma(Y))=\supp_\Gamma(Y)\setminus\{d\}$.
\item Let $\mathcal P_d(\Gamma)=\{Y\in\mathcal P(\Gamma):d\in\supp_\Gamma(Y)\}$. Then
$\Gcirc_{L(\Gamma),d}$ is a quantum access structure if and only if $\mathcal P_d(\Gamma)\neq\emptyset$ and
$\big(\supp_\Gamma(Y)\cap\supp_\Gamma(Z)\big)\setminus\{d\}\neq\emptyset$ for all
$Y,Z\in\mathcal P_d(\Gamma)$.
\item $L(\Gamma)$ is secret-sharing if and only if the condition of (3) holds for every $d\in[n]$.
\end{enumerate}
\end{theorem}
\begin{IEEEproof}
(1) The column of $x\in[n]$ is the unit vector $\e_x$, and the column of $y^*$ is $A_{\cdot,y}=\one_{N(y)}$; so for every $Y\subseteq[n]$, $\sum_{y\in Y}A_{\cdot,y}=A\one_Y=\one_{\mathcal{O}(Y)}$.
Let $X\sqcup Y^*$ be admissible, so $X\cap Y=\emptyset$. A linear relation over $\F_2$ among its columns
with nonzero coefficients exactly on $X'\sqcup Y'^*\subseteq X\sqcup Y^*$ reads $\one_{X'}=\one_{\mathcal{O}(Y')}$,
that is, $X'=\mathcal{O}(Y')$; here $Y'\neq\emptyset$ because the unit vectors are independent, and
$Y'\cap \mathcal{O}(Y')\subseteq Y\cap X=\emptyset$, so $Y'$ is Eulerian. Conversely, for Eulerian $Y'$ the
columns of $C_\Gamma(Y')$ sum to zero. Hence an admissible set is dependent if and only if it contains
$C_\Gamma(Y')$ for some Eulerian $Y'$, and the circuits are the minimal sets $C_\Gamma(Y)$, $Y$ Eulerian.
For Eulerian $Y$ and nonempty $Z\subseteq Y$, we have $C_\Gamma(Z)\subseteq C_\Gamma(Y)$ if and only if
$\mathcal{O}(Z)\subseteq \mathcal{O}(Y)$, and then $Z\cap \mathcal{O}(Z)\subseteq Y\cap \mathcal{O}(Y)=\emptyset$, so $Z$ is Eulerian. Thus $C_\Gamma(Y)$ is minimal exactly when $Y$ is primitive.

(2) The first two statements hold because $C_\Gamma(Y)=\mathcal{O}(Y)\sqcup Y^*$, and $Y\cap \mathcal{O}(Y)=\emptyset$ makes them mutually exclusive. Moreover $\varphi(C_\Gamma(Y))=\supp_\Gamma(Y)$.

(3) and (4) follow from (1), (2), Proposition~\ref{prop:nocloning} and
Corollary~\ref{cor:secret-sharing}.
\end{IEEEproof}

For every vertex $u$, the singleton $\{u\}$ is primitive Eulerian, because $\Gamma$ has no loops; it gives
the circuit $C_\Gamma(\{u\})=N(u)\sqcup\{u^*\}$. This yields a graph condition of Sarvepalli
\cite[Thm.~5]{Sarvepalli2011}.

\begin{corollary}\label{cor:girth5}
If $\Gamma$ has at least one vertex, no cycle of length at most $4$, and no vertex of degree $1$, then $\Gcirc_{L(\Gamma),d}$ is
a quantum access structure for no $d\in[n]$; in particular $L(\Gamma)$ is not secret-sharing.
\end{corollary}
\begin{IEEEproof}
If $d$ is isolated, then $C_\Gamma(\{d\})=\{d^*\}$ and $B_d=\emptyset$. Otherwise $d$ has two neighbours
$u\neq w$. The circuits $C_\Gamma(\{u\})$ and $C_\Gamma(\{w\})$ contain $d$, and their player supports
$(\{u\}\cup N(u))\setminus\{d\}$ and $(\{w\}\cup N(w))\setminus\{d\}$ are disjoint: $u$ and $w$ are not
adjacent, since $\Gamma$ has no triangle, and a common neighbour other than $d$ would give a $4$-cycle.
In both cases Theorem~\ref{thm:graph-circuits}(3) fails.
\end{IEEEproof}

\begin{example}[A path with the dealer in the middle]\label{ex:path}
Let $\Gamma$ be the path $1\,2\,3$ over $\F_2$ and $d=2$. The primitive Eulerian sets are $\{1\}$,
$\{2\}$, $\{3\}$ and $\{1,3\}$. The circuits meeting $J_2$ are those of $\{1\}$, $\{3\}$, $\{2\}$, namely
$\{2,1^*\}$, $\{2,3^*\}$, $\{1,3,2^*\}$, while $\{1^*,3^*\}$ does not meet $J_2$; so
$\Gcirc_{L(\Gamma),2}=\{\{1\},\{3\},\{1,3\}\}$, which is not a quantum access structure. The code
obtained below from the same graph state is nevertheless meaningful: by Corollary~\ref{cor:cutrank},
$\{1,3\}$ is authorized and each of $\{1\}$, $\{3\}$ is intermediate.
\end{example}

Whether $\Gcirc_{L,d}$ is a quantum access structure may depend on the dealer. Among represented
matroids with a circuit through every dealer, self-duality is sufficient for the secret-sharing
property, but it is not necessary.

\begin{example}\label{ex:dealer-dependence}
(1) Let $\Gamma$ be the triangle $1\,2\,3$ with a pendant edge $34$. For $d=1$ the minimal sets
$B_d(C)$ are $\{2,3\}$ and $\{2,4\}$, so $\Gcirc_{L(\Gamma),1}$ is a quantum access structure, and
likewise for $d=2$. For $d=3$ they include $\{1,2\}$ and $\{4\}$, and for $d=4$ they include $\{1,2\}$
and $\{3\}$; so these dealers fail.

(2) Let $\Gamma=K_3$. Here $A^2\neq I$ over $\F_2$, since all degrees are even, so $L(K_3)$ is not
self-dual by Proposition~\ref{prop:graph-selfdual} below. Its primitive Eulerian sets are the three
singletons and $\{1,2,3\}$, with circuits $C_\Gamma(\{u\})=N(u)\sqcup\{u^*\}$ and $\{1^*,2^*,3^*\}$, so for
every dealer $d$ the only member of $\Gmin_{L,d}$ is $P_d$. Hence $L(K_3)$ is secret-sharing. Sarvepalli's
six-qubit example \cite[Sec.~IV]{Sarvepalli2011}, which realizes the $((3,5))$ threshold structure, is
another matroid that is not self-dual but induces a quantum access structure.
\end{example}

\subsection{The dealer-shortened stabilizer code}\label{subsec:dealer}

Let $W=\row(M)\le\Fq^{2n}$ be Lagrangian, and let $d$ be a dealer. By \eqref{eq:split},
$\Fq^{2n}=E_d\perp E_{P_d}$, where $E_d=E_{\{d\}}$ carries the form $\omega_d=\omega|_{E_d}$; thus
\begin{equation}\label{eq:dealer-split}
 \omega\big((a,p),(b,p')\big)=\omega_d(a,b)+\omega(p,p')
\end{equation}
for $a,b\in E_d$ and $p,p'\in E_{P_d}$.
We identify $E_{P_d}$ with $\Fq^{2(n-1)}$. Let $M_d=M[J_d]\in\Fq^{n\times2}$ be the dealer block of
$M$, $M_{P_d}=M[J_{P_d}]$, and $t_d=\rk M_d=\dim_{\Fq}\pi_d(W)$.

\begin{definition}[Dealer shortening]\label{def:shortening}
The \emph{shortening} of $W$ at $d$ is $S_d=\{p\in E_{P_d}:(0,p)\in W\}$, and the stabilizer code of $S_d$ on the players is the \emph{dealer-shortened code} $Q_d(W)$.
\end{definition}

Thus we keep the vectors of $W$ that are zero on the dealer and then delete the dealer coordinates. This is shortening and not puncturing, in the sense used for the CSS construction of
\cite{SarvepalliRaussendorf2010}; puncturing would keep the player parts of all vectors of $W$.

\begin{proposition}\label{prop:shortening}
$t_d\in\{1,2\}$, the space $S_d$ is isotropic of dimension $n-t_d$, and $Q_d(W)$ is an
$[[n-1,t_d-1]]_q$ code. In particular, $Q_d(W)$ encodes one qudit if and only if $\pi_d(W)=E_d$.
\end{proposition}
\begin{IEEEproof}
If $t_d=0$, then $W\le E_{P_d}$ would be isotropic of dimension $n$ in a symplectic space of dimension
$2(n-1)$, which is impossible. The map $p\mapsto(0,p)$ is an isomorphism from $S_d$ onto the kernel
of $\pi_d$ on $W$, so $\dim_{\Fq} S_d=n-t_d$. For $p,p'\in S_d$, \eqref{eq:dealer-split} gives
$\omega(p,p')=\omega((0,p),(0,p'))=0$.
\end{IEEEproof}

We call the dealer $d$ \emph{surjective} if $\pi_d(W)=E_d$; by Proposition~\ref{prop:shortening} this
means exactly that $t_d=\rk M_d=2$, and it is the case in which $Q_d(W)$ encodes one qudit. The stabilizer matrix of $Q_d(W)$ is obtained from
$M$ by elimination. Let $K_d$ be a matrix whose rows form a basis of the left kernel
$\{\lambda\in\Fq^{1\times n}:\lambda M_d=0\}$.

\begin{theorem}[Stabilizer matrix]\label{thm:Hd}
The matrix $H_d=K_dM_{P_d}$ has full row rank $n-t_d$, its row space is $S_d$, and
$H_d\Omega_{n-1}H_d^T=0$. In particular, if $t_d=2$, then $H_d$ is an
$(n-2)\times2(n-1)$ stabilizer matrix of an $[[n-1,1,\delta_d]]_q$ code, where
$\delta_d=d(S_d)$.
\end{theorem}
\begin{IEEEproof}
For a row $\lambda$ of $K_d$, the vector $\lambda M$ lies in $W$ and has dealer part $\lambda M_d=0$,
so its player part $\lambda M_{P_d}$ lies in $S_d$. If $\mu H_d=0$, then $\mu K_dM=0$, because also
$\mu K_dM_d=0$; as $M$ has full row rank, $\mu K_d=0$ and $\mu=0$. Hence $\rk H_d=n-t_d=\dim_{\Fq} S_d$ and
$\row(H_d)=S_d$. The rows of $[\,0\mid H_d\,]=K_dM$ (in the dealer--player order) lie in the isotropic
space $W$, which gives $H_d\Omega_{n-1}H_d^T=0$.
\end{IEEEproof}

When $t_d=2$, a particular choice of $K_d$ also produces the logical operators.

\begin{construction}[Dealer shortening in row-reduced form]\label{con:dealer}
Let $M$ be a generator matrix and consider a surjective dealer $d$, that is \(t_{d}=2\). Choose $R_d\in GL_n(\Fq)$ with $R_dM_d=\left[\begin{smallmatrix}I_2\\\mathbf{0}\end{smallmatrix}\right]$, and order the columns as
$(d,P_d\mid d^*,P_d^*)$, so that
\[ R_dM=\left[\begin{array}{cc|cc}
 1&\alpha_X&0&\alpha_Z\\
 0&\beta_X&1&\beta_Z\\
 \mathbf{0}&H_X&\mathbf{0}&H_Z
 \end{array}\right], \]
with rows $\alpha_X,\alpha_Z,\beta_X,\beta_Z\in\Fq^{1\times(n-1)}$ and
$H_X,H_Z\in\Fq^{(n-2)\times(n-1)}$. Output: the stabilizer matrix $H_d=[\,H_X\mid H_Z\,]$ and the
logical operators $\ell_X=(\alpha_X\mid\alpha_Z)$ and $\ell_Z=-(\beta_X\mid\beta_Z)$.
\end{construction}

Taking $K_d=[\,\mathbf{0}\ \ I_{n-2}\,]R_d$ in Theorem~\ref{thm:Hd} gives exactly this $H_d$. The isotropy of
the rows of $R_dM$ and \eqref{eq:dealer-split} show that $\ell_X$ and $\ell_Z$ are orthogonal to the rows
of $H_d$ and that $\omega(\ell_X,\ell_Z)=1$. Indeed, the first two rows of $R_dM$ are
$(\e_d,\ell_X)$ and $(\e_{d^*},-\ell_Z)$ in the notation $(a,p)$, and
$0=\omega_d(\e_d,\e_{d^*})+\omega(\ell_X,-\ell_Z)=1-\omega(\ell_X,\ell_Z)$. So the rows of $H_d$ followed
by $\ell_X,\ell_Z$ have the block-diagonal Gram matrix
$\mathrm{diag}\big(\mathbf{0}_{n-2},\left(\begin{smallmatrix}0&1\\-1&0\end{smallmatrix}\right)\big)$, the
standard form of a stabilizer matrix together with one pair of logical operators. If $W$ is a CSS subspace, this is the
block matrix of \cite{SarvepalliRaussendorf2010}; see Proposition~\ref{prop:css}.

For a surjective dealer let $\cL_d=S_d^\perp/S_d$, the logical space of $Q_d(W)$, with the induced form
$\bar\omega([p],[p'])=\omega(p,p')$, and write $[p]=p+S_d$. Its dimension is $2$.

\begin{theorem}\label{thm:Theta}
Let $d$ be surjective. For $a\in E_d$ choose $p\in E_{P_d}$ with $(a,p)\in W$ and put $\Theta_d(a)=[p]$. Then for \(a,b\in E_d\), $\Theta_d:E_d\to\cL_d$ is a well-defined linear isomorphism with
\[ \bar\omega\big(\Theta_d(a),\Theta_d(b)\big)=-\omega_d(a,b). \]
In Construction~\ref{con:dealer}, $\Theta_d(\e_d)=[\ell_X]$ and $\Theta_d(\e_{d^*})=-[\ell_Z]$.
\end{theorem}
\begin{IEEEproof}
A lift $(a,p)\in W$ exists because $\pi_d$ is onto. For $s\in S_d$, $\omega(p,s)=\omega((a,p),(0,s))=0$
by \eqref{eq:dealer-split}, so $p\in S_d^\perp$. Two lifts of $a$ differ by $(0,p-p')\in W$, so
$p-p'\in S_d$, and $\Theta_d$ is well defined; it is linear because lifts can be added and scaled. For
lifts $(a,p)$ and $(b,p')$, isotropy of $W$ and \eqref{eq:dealer-split} give
$0=\omega_d(a,b)+\omega(p,p')$. If $\Theta_d(a)=0$, then $\omega_d(a,b)=0$ for every $b$, so $a=0$;
since both spaces have dimension $2$, $\Theta_d$ is an isomorphism. The first two rows of $R_dM$ are
lifts of $\e_d$ and $\e_{d^*}$.
\end{IEEEproof}

In characteristic $2$ the map $\Theta_d$ is both symplectic and anti-symplectic. In odd characteristic the canonical map $\Theta_d$ reverses the form, and this is the reason for the sign of $\ell_Z$ in
Construction~\ref{con:dealer}.

For $A\subseteq P_d$ let
\begin{align*} D_d(A)&=\pi_d\big(W\cap(E_d\oplus E_A)\big)\\ &=\{a\in E_d:\ (a,p)\in W\text{ for some }p\in E_A\}, \end{align*}
and let $\cL_d(A)=\cL_{S_d}(A)$ be the space of logical classes of $Q_d(W)$ with a representative
supported on $A$. Note that $D_d(A)\subseteq D_d(A')$ if $A\subseteq A'$.

\begin{proposition}\label{prop:Dmatrix}
Let $A\subseteq P_d$ and $\bar A=P_d\setminus A$, and write $\rho=\rho_M$. Then
\begin{gather*} D_d(A)=\{\lambda M_d:\ \lambda\in\Fq^{1\times n},\ \lambda M[J_{\bar A}]=0\},\\
 \dim_{\Fq} D_d(A)=\rho(\bar A\cup\{d\})-\rho(\bar A). \end{gather*}
If $d$ is surjective, then $\Theta_d(D_d(A))=\cL_d(A)$; in particular $\dim_{\Fq} D_d(A)=\ell_{S_d}(A)$.
\end{proposition}
\begin{IEEEproof}
Every $w\in W$ is $\lambda M$ for a unique $\lambda$, and $w\in E_d\oplus E_A$ if and only if
$\lambda M[J_{\bar A}]=0$; then $\pi_d(w)=\lambda M_d$. The space of such $\lambda$ has dimension
$n-\rho(\bar A)$, and the kernel of $\lambda\mapsto\lambda M_d$ on it has dimension
$n-\rho(\bar A\cup\{d\})$. If $a\in D_d(A)$ with lift $(a,p)$, $p\in E_A$, then $\Theta_d(a)=[p]$ lies
in $\cL_d(A)$. Conversely, if $\Theta_d(a)=[p']$ with $p'\in S_d^\perp\cap E_A$ and $(a,p)\in W$, then
$(0,p-p')\in W$, so $(a,p')\in W\cap(E_d\oplus E_A)$ and $a\in D_d(A)$.
\end{IEEEproof}

By Construction~\ref{con:dealer}, the symplectic matroid of $S_d$ has as
bases the admissible $(n-2)$-subsets $B'$ of $J_{P_d}$ with $\rk M[J_d\cup B']=n$, since
$R_dM[J_d\cup B']=\left[\begin{smallmatrix}I_2&*\\\mathbf{0}&H_d[B']\end{smallmatrix}\right]$. These are ranks of
inadmissible sets, so they are in general not determined by $L$; even $t_d=\rho(\{d\})$ is not
(Example~\ref{ex:q3}).

\subsection{Encoding, recovery and the stabilizer state}\label{subsec:encoding}

\begin{theorem}\label{thm:encoding}
Let $d$ be surjective, and let $\ell_X,\ell_Z$ be as in Construction~\ref{con:dealer}.
\begin{enumerate}
\item $S_d^{(0)}=S_d+\Fq\ell_Z$ is a Lagrangian subspace of $E_{P_d}$. Choose a lift of $S_d^{(0)}$ as in
Lemma~\ref{lem:lift}; its restriction to $S_d$ is a lift of $S_d$, and we realize $Q_d(W)$ as the code of
this restricted lift. Let $|\bar0\rangle$ be the stabilizer state of the lift of $S_d^{(0)}$, and put
$|\bar a\rangle=D(a\ell_X)|\bar0\rangle$ for $a\in\Fq$. Then $\{|\bar a\rangle:a\in\Fq\}$ is an
orthonormal basis of $Q_d(W)$, so $V_d|a\rangle=|\bar a\rangle$ defines an encoding isometry
$V_d \colon \C^q\to Q_d(W)$. Moreover $\Dh(b\ell_Z)|\bar a\rangle=\chi(ab)|\bar a\rangle$, and
$D(b\ell_X)|\bar a\rangle$ is a multiple of $|\overline{a+b}\rangle$ by a phase, for all $a,b\in\Fq$.
\item Let $A\subseteq P_d$ with $D_d(A)=E_d$, and fix $a_0\in A$. There are
$x_A,z_A\in S_d^\perp\cap E_A$ with $[x_A]=[\ell_X]$ and $[z_A]=[\ell_Z]$, and there is a Clifford unitary
$U_A$ on the qudits in $A$ such that $U_AD(bx_A)U_A^\dagger$ is a multiple of $X_{a_0}(b)$ and
$U_AD(bz_A)U_A^\dagger$ is a multiple of $Z_{a_0}(b)$ for all $b\in\Fq$; here $X_{a_0}(b)$ and
$Z_{a_0}(b)$ act on qudit $a_0$ only. For every such $U_A$ there are a unit vector $|\eta\rangle$ of the
qudits in $P_d\setminus\{a_0\}$ and a unitary $U_0$ on $\C^q$ with
\[ (U_A\otimes I)V_d|\psi\rangle=U_0|\psi\rangle_{a_0}\otimes|\eta\rangle, \qquad |\psi\rangle\in\C^q. \]
Consequently $\mathcal R_A(\varrho_A)=U_0^\dagger\,\Tr_{A\setminus\{a_0\}}\big(U_A\varrho_AU_A^\dagger\big)U_0$
is a recovery channel for $A$.
\end{enumerate}
\end{theorem}
\begin{IEEEproof}
(1) By Theorem~\ref{thm:Theta}, $[\ell_Z]\neq0$, so $\ell_Z\notin S_d$ and $\dim_{\Fq}S_d^{(0)}=n-1$, half
the dimension of $E_{P_d}$. It is isotropic because $\ell_Z\in S_d^\perp$ and $\omega(\ell_Z,\ell_Z)=0$.
A lift of $S_d^{(0)}$ restricts to a homomorphism on $S_d$ with the properties of Lemma~\ref{lem:lift}.
Since $\ell_X\in S_d^\perp$, the operator $D(a\ell_X)$ commutes with $\Dh(s)$ for every $s\in S_d$, so
$|\bar a\rangle\in Q_d(W)$. By \eqref{eq:commutation} and $\omega(\ell_X,\ell_Z)=1$,
\begin{align*} \Dh(b\ell_Z)D(a\ell_X)&=\chi\big(\omega(a\ell_X,b\ell_Z)\big)D(a\ell_X)\Dh(b\ell_Z)\\
 &=\chi(ab)\,D(a\ell_X)\Dh(b\ell_Z), \end{align*}
and $\Dh(b\ell_Z)|\bar0\rangle=|\bar0\rangle$; this gives the eigenvalue $\chi(ab)$. If $a\neq a'$, the
characters $b\mapsto\chi(ab)$ and $b\mapsto\chi(a'b)$ differ, because $\chi$ is nontrivial; so
$|\bar a\rangle$ and $|\bar a'\rangle$ are eigenvectors of some unitary $\Dh(b\ell_Z)$ with different
eigenvalues, hence orthogonal. They are unit vectors, there are $q=\dim_{\C}Q_d(W)$ of them, and so they
form an orthonormal basis. Finally $D(b\ell_X)D(a\ell_X)$ is a multiple of $D((a+b)\ell_X)$ by a phase,
by the multiplication rule $X(a)Z(b)X(a')Z(b')=\chi(b\cdot a')X(a+a')Z(b+b')$.

(2) By Proposition~\ref{prop:Dmatrix}, $\cL_d(A)=\Theta_d(E_d)=\cL_d$, which gives $x_A$ and $z_A$; and
$\omega(x_A,z_A)=\bar\omega([\ell_X],[\ell_Z])=1$. The space $E_A$ is nondegenerate by \eqref{eq:split},
and $\omega(\e_{a_0},\e_{a_0^*})=1$, where $\e_{a_0}$ and $\e_{a_0^*}$ are the unit vectors of the
coordinates $a_0$ and $a_0^*$. By Lemma~\ref{lem:symplectic-basis} there is an $\Fq$-linear map
$F:E_A\to E_A$ preserving $\omega$ with $F(x_A)=\e_{a_0}$ and $F(z_A)=\e_{a_0^*}$. It preserves
$\tr_{\mathbb{F}_{q}/\mathbb{F}_{p}}\circ\omega$, so it is induced by a Clifford unitary $U_A$, in the sense that $U_AD(v)U_A^\dagger$ is
a multiple of $D(Fv)$ for every $v\in E_A$ \cite{HostensDehaeneDeMoor2005}; to apply this, identify
$\C^q$ with $(\C^p)^{\otimes m}$ by expanding $x\in\Fq$ in a basis of $\Fq$ over $\F_p$ and $b\in\Fq$ in
the trace-dual basis, which turns $\tr_{\mathbb{F}_{q}/\mathbb{F}_{p}}\circ\omega$ into the standard symplectic form over $\F_p$ and each
$D(v)$ into a tensor product of $p$-dimensional Pauli operators. Since $F$ is $\Fq$-linear,
$F(bx_A)=b\,\e_{a_0}$ and $F(bz_A)=b\,\e_{a_0^*}$, and $D(b\,\e_{a_0})=X_{a_0}(b)$,
$D(b\,\e_{a_0^*})=Z_{a_0}(b)$.

Let $Q'=(U_A\otimes I)Q_d(W)$. The operators $D(bx_A)$ and $D(bz_A)$ map $Q_d(W)$ to itself, because
$bx_A,bz_A\in S_d^\perp$; hence $X_{a_0}(b)$ and $Z_{a_0}(b)$ map $Q'$ to itself. The products
$X_{a_0}(b)Z_{a_0}(c)$, $b,c\in\Fq$, span the full matrix algebra $\mathcal M_q(\C)$ of qudit $a_0$, so
$Q'$ is invariant under $\mathcal M_q(\C)\otimes I$. Every such subspace has the form $\C^q\otimes\mathcal K$
for a subspace $\mathcal K$ of the other qudits (apply the matrix units $|x\rangle\langle y|\otimes I$),
and $\dim_{\C}Q'=q$ forces $\mathcal K=\C|\eta\rangle$ for a unit vector $|\eta\rangle$. Then
$(U_A\otimes I)V_d$ is an isometry of $\C^q$ onto $\C^q\otimes|\eta\rangle$, which gives $U_0$. Finally,
$U_A$ acts only on $A$, so $\Tr_{A\setminus\{a_0\}}(U_A\varrho_AU_A^\dagger)$ is the reduced state on
$a_0$ of
$(U_A\otimes I)V_d\varrho V_d^\dagger(U_A\otimes I)^\dagger=U_0\varrho U_0^\dagger\otimes|\eta\rangle\langle\eta|$,
that is, $U_0\varrho U_0^\dagger$.
\end{IEEEproof}

The recovery map in Theorem~\ref{thm:encoding} can be obtained by solving for $x_A,z_A$ by elimination, extending them to a symplectic basis, and choosing a Clifford implementing the resulting symplectic map. To fix the remaining unitary, set
$|\Psi_a\rangle=(U_A\otimes I)V_d|a\rangle$ and choose a unit vector $|\eta\rangle$ spanning the image of $\Tr_{a_0}|\Psi_0\rangle\langle\Psi_0|$. Then
$(U_0)_{b,a}=(\langle b|_{a_0}\otimes\langle\eta|)|\Psi_a\rangle$ gives its matrix entries.
The proof shows that this matrix is unitary, so these choices determine the recovery channel.

The code $Q_d(W)$ is also the code that the stabilizer state of $W$ implements; this is how graph states
are used for secret sharing \cite{MarkhamSanders2008}. In the next proposition the dealer is written as
the first tensor factor, $(\C^q)^{\otimes n}=\C^q\otimes(\C^q)^{\otimes P_d}$, and
$|\Phi\rangle=q^{-1/2}\sum_{x\in\Fq}|x\rangle|x\rangle\in\C^q\otimes\C^q$.

\begin{proposition}[Stabilizer state and encoding]\label{prop:choi}
Let $\Dh$ be a lift of $W$ as in Lemma~\ref{lem:lift} and $|\psi_W\rangle$ its stabilizer state, a unit
vector spanning the image of $\Pi_W$.
\begin{enumerate}
\item For $Y\subseteq[n]$, the reduced state $\varrho_Y$ of $|\psi_W\rangle$ is the maximally mixed state on
a subspace $\mathcal H_Y$ with $\dim_{\C}\mathcal H_Y=q^{\rho_M(Y)-|Y|}$; thus its von Neumann entropy is
$(\rho_M(Y)-|Y|)\log q$.
\end{enumerate}
Assume that $d$ is surjective.
\begin{enumerate}
\setcounter{enumi}{1}
\item $\varrho_d=I/q$. For $p\in S_d$ the operator $\Dh(0,p)$ equals $I\otimes g(p)$ for a unique operator
$g(p)$ on the players, the map $p\mapsto g(p)$ is a lift of $S_d$, and the support of $\varrho_{P_d}$ is
the code space $Q$ of this lift, a realization of $Q_d(W)$.
\item There is an isometry $V \colon \C^q\to(\C^q)^{\otimes P_d}$ with image $Q$ such that
$|\psi_W\rangle=(I\otimes V)|\Phi\rangle$.
\item Let $a=(\alpha\mid\beta)\in E_d$, with $\alpha,\beta\in\Fq$, and let $(a,p)\in W$. Then
\[ D(p)\,V=c\,V\,X(\alpha)Z(-\beta) \]
for some $c\in\mu$. Here $Z(-\beta)=Z(\beta)^{-1}=Z(\beta)^\dagger$; in characteristic $2$,
$X(\alpha)Z(-\beta)=X(\alpha)Z(\beta)$.
\end{enumerate}
\end{proposition}
\begin{IEEEproof}
(1) Since $|\psi_W\rangle\langle\psi_W|=\Pi_W=q^{-n}\sum_{w\in W}\Dh(w)$, and the partial trace over the
qudits outside $Y$ of a Pauli operator with a nontrivial factor there is zero,
$\varrho_Y=q^{-|Y|}\sum_{w\in W\cap E_Y}\Dh(w)|_Y$, where $\Dh(w)|_Y$ is the factor of $\Dh(w)$ on $Y$.
These factors form a lift of the isotropic space $W\cap E_Y$, viewed in $E_Y$, so by
Lemma~\ref{lem:lift} $\varrho_Y$ is $q^{s-|Y|}$ times a projection of rank $q^{|Y|-s}$, where
$s=\dim_{\Fq}(W\cap E_Y)=2|Y|-\rho_M(Y)$ by Lemma~\ref{lem:counting}.

(2) By (1), $\varrho_d$ is maximally mixed on a space of dimension $q^{\rho_M(\{d\})-1}=q^{t_d-1}=q$,
that is, $\varrho_d=I/q$. If $p\in S_d$, then $(0,p)\in W$ and $\Dh(0,p)$ is a multiple of
$I\otimes D(p)$; write it as $I\otimes g(p)$. The map $g$ is a homomorphism because $\Dh$ is, it takes
values in multiples of $D(p)$ by elements of $\mu$, and $g(p)$ is a scalar only if $p=0$; so it is a lift
of $S_d$. Since $(I\otimes g(p))|\psi_W\rangle=|\psi_W\rangle$, every vector
$(\langle\phi|\otimes I)|\psi_W\rangle$ is fixed by all $g(p)$, and these vectors span the support of
$\varrho_{P_d}$. Hence the support lies in $Q$. By the Schmidt decomposition the support has dimension
$\rk\varrho_d=q$, and $\dim_{\C}Q=q$ by Lemma~\ref{lem:lift}, so they are equal.

(3) Write $|\psi_W\rangle=\sum_{x\in\Fq}|x\rangle\otimes|\phi_x\rangle$, where
$|\phi_x\rangle=(\langle x|\otimes I)|\psi_W\rangle$.
The equality $\varrho_d=I/q$ says $\langle\phi_y|\phi_x\rangle=\delta_{xy}/q$.
Thus $V|x\rangle=q^{1/2}|\phi_x\rangle$ defines an isometry with image $Q$ by (2), and
$(I\otimes V)|\Phi\rangle=|\psi_W\rangle$ by substitution.

(4) Since $(a,p)\in W$, the operator $\Dh(a,p)=c'\,K\otimes D(p)$ with $K=X(\alpha)Z(\beta)$ and some
$c'\in\mu$ fixes $|\psi_W\rangle$. For every operator $K$ on $\C^q$ one has
$(K\otimes I)|\Phi\rangle=(I\otimes K^T)|\Phi\rangle$, the transpose being taken in the basis
$\{|x\rangle\}$. Hence
\begin{align*}
 (I\otimes V)|\Phi\rangle&=c'(K\otimes D(p))(I\otimes V)|\Phi\rangle\\
 &=c'\big(I\otimes D(p)VK^T\big)|\Phi\rangle.
\end{align*}
If $T$ is an operator on $\C^q$ with values in any space and $(I\otimes T)|\Phi\rangle=0$, then
$T|x\rangle=0$ for all $x$, so $T=0$. Therefore $V=c'D(p)VK^T$, that is,
$D(p)V=c'^{-1}V(K^T)^{-1}$. Now $Z(\beta)$ is diagonal and $X(\alpha)$ is the permutation matrix of
$x\mapsto x+\alpha$, so $K^T=Z(\beta)X(\alpha)^T=Z(\beta)X(-\alpha)$ and
$(K^T)^{-1}=X(\alpha)Z(\beta)^{-1}=X(\alpha)Z(-\beta)$. Put $c=c'^{-1}$.
\end{IEEEproof}

Part (4) is the operator form of Theorem~\ref{thm:Theta}: the logical operator $D(p)$, whose class is
$\Theta_d(a)$, acts on the encoded secret as the Pauli operator of $(\alpha\mid-\beta)$, and the map
$(\alpha\mid\beta)\mapsto(\alpha\mid-\beta)$ reverses $\omega_d$. Hence, if the dealer performs a generalized Bell
measurement on the secret and on the dealer qudit of $|\psi_W\rangle$, the players hold the secret
encoded by $V$ up to a Pauli operator on the secret that is determined by the outcome
\cite{BennettEtAl1993,Werner2001}. The outcome is uniformly distributed and independent of the secret,
so it may be announced, and the access structure of this protocol is that of $Q_d(W)$. The isometries
$V$ and $V_d$ map onto realizations of $Q_d(W)$ for two lifts of $S_d$, which differ at most by a Pauli
operator on the players (Section~\ref{subsec:stabilizer}); so $V$ and $V_d$ have the same authorized,
forbidden and intermediate sets.

\subsection{The operational access structure}\label{subsec:operational}

For a surjective dealer $d$ we write $\Grec_d(W)$, $\Gforb_d(W)$ and $\Gint_d(W)$ for the families of
authorized, forbidden and intermediate player sets of $Q_d(W)$, and $\rho=\rho_M$.

\begin{theorem}\label{thm:operational}
Let $d$ be surjective, $A\subseteq P_d$ and $\bar A=P_d\setminus A$.
\begin{enumerate}
\item $\dim_{\Fq} D_d(A)=\rho(\bar A\cup\{d\})-\rho(\bar A)=2+\rho(A)-\rho(A\cup\{d\})=I(A:d)_{\psi_W}/\log q$.
\item $\dim_{\Fq} D_d(A)+\dim_{\Fq} D_d(\bar A)=2$, and $D_d(\bar A)$ is the $\omega_d$-orthogonal complement of
$D_d(A)$ in $E_d$.
\item $A\in\Grec_d(W)$ if and only if $D_d(A)=E_d$; $A\in\Gforb_d(W)$ if and only if $D_d(A)=0$; and
$A\in\Gint_d(W)$ if and only if $D_d(A)$ is a line $\ell$. In the last case $D_d(\bar A)=\ell$, and the
logical classes supported on $A$, as well as those supported on $\bar A$, are exactly $\Theta_d(\ell)$;
the corresponding logical operators commute.
\end{enumerate}
In particular $A\in\Grec_d(W)$ if and only if $\bar A\in\Gforb_d(W)$, and $\Gint_d(W)$ is closed under
complementation.
\end{theorem}
\begin{IEEEproof}
(1) The first expression is Proposition~\ref{prop:Dmatrix}. The last statement of
Lemma~\ref{lem:counting}, applied to $A$ and to $A\cup\{d\}$, whose complements in $[n]$ are
$\bar A\cup\{d\}$ and $\bar A$, gives $\rho(\bar A\cup\{d\})=n-2|A|+\rho(A)$ and
$\rho(\bar A)=n-2|A|-2+\rho(A\cup\{d\})$, hence the second expression. By
Proposition~\ref{prop:choi}(1), $I(A:d)=\big(\rho(A)+\rho(\{d\})-\rho(A\cup\{d\})\big)\log q$, and
$\rho(\{d\})=t_d=2$.

(2) By Proposition~\ref{prop:Dmatrix}, $\Theta_d$ maps $D_d(A)$ and $D_d(\bar A)$ onto $\cL_d(A)$ and
$\cL_d(\bar A)$, and these are orthogonal complements by Lemma~\ref{lem:logical}(1) with $k=1$. Since
$\Theta_d$ reverses the form, it maps orthogonal complements to orthogonal complements.

(3) By Lemma~\ref{lem:erasure}, $A$ is authorized if and only if no nontrivial logical class is
supported on $\bar A$, that is, $D_d(\bar A)=0$, or by (2), $D_d(A)=E_d$. Likewise $A$ is forbidden if
and only if $D_d(A)=0$. If $D_d(A)=\ell$ is a line, then $D_d(\bar A)=\ell^\perp=\ell$, because a line in
the two-dimensional symplectic space $E_d$ is its own orthogonal complement, and its image under
$\Theta_d$ is isotropic.
\end{IEEEproof}

Statement (1) identifies the pair-rank criterion with the information-theoretic description of quantum secret sharing \cite{ImaiEtAl2005}, in which a set is authorized when its mutual information with a reference system purifying the secret is maximal and forbidden when it becomes zero; by Proposition~\ref{prop:choi}, the dealer qudit of $|\psi_W\rangle$ is such a reference system. The scheme
$V_d$ is perfect exactly when $\Gint_d(W)=\emptyset$; otherwise it is a ramp scheme. An intermediate set $A$ has $I(A:d)=\log q$, half of the maximum $2\log q$, and what it can learn about the secret is carried by the commuting logical operators with classes in $\Theta_d(\ell)$.

\begin{corollary}\label{cor:cutrank}
Let $M=[I_n\mid A]$ with $A$ symmetric over $\Fq$. Then $d$ is surjective if and only if $d$ has a
neighbour, and in that case, for $A'\subseteq P_d$,
\[ \dim_{\Fq} D_d(A')=1+\cutrk_A(A')-\cutrk_A(A'\cup\{d\}). \]
So $A'$ is authorized if and only if $\cutrk_A(A'\cup\{d\})=\cutrk_A(A')-1$, and intermediate if and
only if the two cut-ranks are equal.
\end{corollary}
\begin{IEEEproof}
By Lemma~\ref{lem:cutrank}, $t_d=\rho(\{d\})=1+\cutrk_A(\{d\})$, and $\cutrk_A(\{d\})=1$ if and only if
some $A_{dj}$, $j\neq d$, is nonzero. Insert Lemma~\ref{lem:cutrank} into the second expression of
Theorem~\ref{thm:operational}(1).
\end{IEEEproof}

\begin{corollary}\label{cor:invariance}
Let $d$ be surjective. The families $\Grec_d(W)$, $\Gforb_d(W)$, $\Gint_d(W)$ and the distance
$\delta_d=\min\{|B|:D_d(B)\neq0\}$ of $Q_d(W)$ depend only on the pair-rank function $\rho_M$. In
particular they do not change if the two columns of each pair are transformed by a matrix in
$SL_2(\Fq)$, that is, under $\Fq$-linear local Clifford operations.
\end{corollary}
\begin{IEEEproof}
By Proposition~\ref{prop:Dmatrix}, a nontrivial logical class of $Q_d(W)$ is supported on $B$ if and
only if $D_d(B)\neq0$, which gives the formula for $\delta_d$, and $\dim_{\Fq} D_d(B)$ is a function of
$\rho_M$ by Theorem~\ref{thm:operational}(1). A matrix in $SL_2(\Fq)$ preserves the form on each
$E_{\{i\}}$ and the column span of each pair.
\end{IEEEproof}

For $q=p$ every local Clifford operation acts on each pair through $SL_2(\F_p)$. For $m>1$ a local
Clifford operation acts on $E_{\{i\}}\cong\F_p^{2m}$ by a map that preserves $\tr_{\mathbb{F}_{q}/\mathbb{F}_{p}}\circ\omega$ but need not
be $\Fq$-linear, so it may take an $\Fq$-linear code to one that is only $\F_p$-linear; the weight
enumerators are still preserved, since such operations preserve supports.

The circuits of $L$ give upper bounds on $\delta_d$.

\begin{lemma}\label{lem:circuit-bound}
Let $W$ represent $L$, let $d$ be surjective and $C\in\cC_d(L)$. Then $D_d(B_d(C))\neq0$, and
$\delta_d\le|C|-1$.
\end{lemma}
\begin{IEEEproof}
The vector $w_C$ of Lemma~\ref{lem:circuit-vector} lies in $W\cap(E_d\oplus E_{B_d(C)})$ and has a
nonzero dealer part, and $|B_d(C)|=|C|-1$.
\end{IEEEproof}

The bound is strict in Theorem~\ref{thm:Gstar}: $\delta_d=1$, whereas the smallest circuit
meeting $J_d$ has $|C|-1=3$. Vectors with support containing both elements of a pair can
therefore give logical operators smaller than every dealer circuit.

The information-group viewpoint \cite{GheorghiuLooiGriffiths2010,Gheorghiu2012} also makes
the nature of intermediate information explicit.

\begin{proposition}\label{prop:intermediate-channel}
Let $A\in\Gint_d(W)$ and $\mathcal N_A(\varrho)=\Tr_{\bar A}(V_d\varrho V_d^\dagger)$.
There are an orthonormal basis $\{|t\rangle:t\in\Fq\}$ of the input and density operators
$\sigma_t^A$ with pairwise orthogonal supports such that
\[ \mathcal N_A(\varrho)=\sum_{t\in\Fq}\langle t|\varrho|t\rangle\sigma_t^A. \]
Thus $A$ recovers one $q$-outcome classical observable and no coherence in this basis.
\end{proposition}
\begin{IEEEproof}
By Theorem~\ref{thm:operational}, $A$ and $\bar A$ support the same logical line. Choose coherent Pauli lifts of this line and label their joint eigenbasis so that its logical
action is $Z(b)|t\rangle=\chi(bt)|t\rangle$, $b\in\Fq$. Such a basis follows from Theorem~\ref{thm:encoding} after choosing this line as the logical $Z$-line. For each $b$, both
$A$ and $\bar A$ have a unitary representative of $Z(b)$, with phases adjusted on the code. Tracing out the representative on $\bar A$ gives $\mathcal N_A(|t\rangle\langle u|)=\chi(b(t-u))\mathcal N_A(|t\rangle\langle u|)$. For $t\ne u$, some $b$ makes this character different from one, so the off-diagonal image is zero. Put $\sigma_t^A=\mathcal N_A(|t\rangle\langle t|)$. Its support lies in the joint eigenspace of the representatives on $A$ with eigencharacter $b\mapsto\chi(bt)$. Distinct eigencharacters give orthogonal supports. Then by linearity we have the formula.
\end{IEEEproof}

Intermediate information can be hidden by distributing a Pauli correction key classically \cite{NascimentoMullerQuadeImai2001,Gheorghiu2012}. The following formulation also realizes a specified subfamily of the operational access structure.

\begin{proposition}\label{prop:hybrid}
Let $d$ be surjective and let $\emptyset\ne\Gamma\subseteq\Grec_d(W)$ be monotone.
Choose a uniform key $K=(\alpha,\beta)\in\Fq^2$, encode
$V X(\alpha)Z(\beta)\varrho Z(\beta)^\dagger X(\alpha)^\dagger V^\dagger$, and share $K$ using a
perfect classical scheme with access structure $\Gamma$ \cite{ItoSaitoNishizeki1987}.
The resulting mixed hybrid scheme is perfect and has exactly access structure $\Gamma$.
Each player receives one qudit and a finite classical share. Equivalently, one can privately share
the uniform Bell outcome in the teleportation protocol after Proposition~\ref{prop:choi}.
\end{proposition}
\begin{IEEEproof}
An authorized $A\in\Gamma$ reconstructs $K$, recovers the encrypted input since
$A\in\Grec_d(W)$, and applies the inverse Pauli operator. If $A\notin\Gamma$, its classical share $s$ has a distribution $p_A(s)$ independent of $K$. The classical sharing randomness is independent of the quantum encoding conditional on $K$. Put
$\tau_A=\sum_s p_A(s)|s\rangle\langle s|$ and $D(K)=X(\alpha)Z(\beta)$. The joint marginal is therefore
\begin{align*}
 \mathcal E_A(\varrho)
 &=\tau_A\otimes\mathcal N_A\Big(q^{-2}\sum_K D(K)\varrho D(K)^\dagger\Big)\\
 &=\tau_A\otimes\mathcal N_A(I/q),
\end{align*}
where $\mathcal N_A=\Tr_{\bar A}\circ(V\,\cdot\,V^\dagger)$ and the second equality is the Pauli
twirl. It is independent of $\varrho$. By linearity the channel is a replacement channel, so
the same privacy holds when the input is entangled with an external reference.

For completeness, a finite classical scheme can be built as follows. For each minimal
$T\in\Gamma$, independently choose $|T|-1$ uniform vectors in $\Fq^2$ and choose the last so
their sum is $K$; give the vector indexed by $i\in T$ to player $i$. A set containing $T$
recovers $K$. A set containing no such $T$ misses at least one additive share in every group,
and its joint classical distribution is independent of $K$. If player $i$ belongs to $m_i$
minimal sets, its classical register has dimension $q^{2m_i}$.
\end{IEEEproof}

\begin{corollary}\label{cor:hybrid-circuit}
If $L$ is self-dual and $\cC_d(L)\ne\emptyset$, then $\Gcirc_{L,d}$ has a perfect hybrid
realization using $Q_d(W)$ and classical shares. This assertion does not require
$\Gcirc_{L,d}=\Grec_d(W)$.
\end{corollary}
\begin{IEEEproof}
By Theorem~\ref{thm:selfdual}, $d$ is surjective and
$\emptyset\ne\Gcirc_{L,d}\subseteq\Grec_d(W)$. Apply Proposition~\ref{prop:hybrid}.
\end{IEEEproof}

The extra classical registers increase share dimensions; this construction does not assert a
pure perfect realization with one qudit per player. The family $\Grec_d(W)$ is a quantum access
structure: $P_d$ reconstructs, and if two reconstructing sets were disjoint, one would be
contained in the forbidden complement of the other, contrary to Theorem~\ref{thm:operational}.

Finally, the CSS construction of Sarvepalli and Raussendorf is the special case in which $W$ is built from a classical code and its dual. Let $\mathsf C\le\Fq^n$ be a linear code with generator matrix $G$, let $\mathsf C^\perp=\{y\in\Fq^n:c\cdot y=0\ \forall c\in\mathsf C\}$ be its dual, with generator matrix $G^\perp$, and let $N$ be the matroid on $[n]$ of the columns $g_1,\dots,g_n$ of $G$; its dual $N^{\diamond}$ is the
matroid of the columns $g^\perp_1,\dots,g^\perp_n$ of $G^\perp$. For $d\in[n]$ let
\begin{align*}
 \mathrm{pu}_d(\mathsf C)&=\{(c_i)_{i\ne d}:c\in\mathsf C\}\le\Fq^{P_d},\\
 \mathrm{sh}_d(\mathsf C)&=\{(c_i)_{i\ne d}:c\in\mathsf C,\ c_d=0\}\le\Fq^{P_d}
\end{align*}
be the \emph{punctured} and the \emph{shortened} code at the coordinate $d$. Then
$\mathrm{sh}_d(\mathsf C^\perp)=\mathrm{pu}_d(\mathsf C)^\perp$, the dual being taken in $\Fq^{P_d}$:
a word $y\in\Fq^{P_d}$, extended by $0$ at $d$, is orthogonal to every $c\in\mathsf C$ exactly when $y$ is
orthogonal to every punctured word. For $A\subseteq[n]$, $\cl_N(A)$ is the closure of $A$ in $N$, that is,
the set of $i$ with $g_i\in\Span_{\Fq}\{g_a:a\in A\}$.

\begin{proposition}\label{prop:css}
Let $W_{\mathsf C}=\{(c\mid c'):c\in\mathsf C,\ c'\in\mathsf C^\perp\}\le\Fq^{2n}$.
\begin{enumerate}
\item $W_{\mathsf C}$ is Lagrangian. Its Lagrangian matroid $L_{\mathsf C}$ has the bases $B\sqcup([n]\setminus B)^*$, $B$ a basis of $N$, and $L_{\mathsf C}$ is self-dual if and only if $N$ is identically self-dual, that is, $N=N^{\diamond}$.
\item $d$ is surjective if and only if $d$ is neither a loop nor a coloop of $N$. In that case $S_d=\{(c\mid c'):c\in\mathrm{sh}_d(\mathsf C),\ c'\in\mathrm{pu}_d(\mathsf C)^\perp\}$, so $S_d$ has the
stabilizer matrix $\left[\begin{smallmatrix}G_1&\mathbf{0}\\\mathbf{0}&G_2\end{smallmatrix}\right]$, where $G_1$ and $G_2$ are generator
matrices of $\mathrm{sh}_d(\mathsf C)$ and $\mathrm{pu}_d(\mathsf C)^\perp$. This is the stabilizer matrix
of \cite{SarvepalliRaussendorf2010}.
\item For $A\subseteq P_d$, $\dim_{\Fq}D_d(A)=[\,d\in\cl_N(A)\,]+[\,d\in\cl_{N^\diamond}(A)\,]$, where $[\cdot]$ is
$1$ if the condition holds and $0$ otherwise.
\item If $N$ is identically self-dual and $d$ is surjective, then $V_d$ is perfect and
$\Grec_d(W_{\mathsf C})=\Gcirc_{L_{\mathsf C},d}=\{A\subseteq P_d:d\in\cl_N(A)\}$.
\end{enumerate}
\end{proposition}
\begin{IEEEproof}
(1) $\dim_{\Fq}W_{\mathsf C}=\dim_{\Fq}\mathsf C+\dim_{\Fq}\mathsf C^\perp=n$, and
$\omega((c\mid0),(0\mid c'))=c\cdot c'=0$, while $\omega$ becomes zero on pairs of vectors of the form
$(c\mid0)$ and on pairs of the form $(0\mid c')$. Let $k_0=\dim_{\Fq}\mathsf C$ and
$M=\left[\begin{smallmatrix}G&\mathbf{0}\\\mathbf{0}&G^\perp\end{smallmatrix}\right]$. For an admissible transversal
$X\sqcup Y^*$ we have $Y=[n]\setminus X$, and $M[X\sqcup Y^*]$ has the blocks $G[X]$ and $G^\perp[Y]$. If
$|X|\neq k_0$, one block has more columns than rows, so the determinant is zero. If $|X|=k_0$, the
determinant is $\pm\det G[X]\det G^\perp[Y]$, which is nonzero if and only if $X$ is a basis of $N$ and $Y$
a basis of $N^\diamond$; the second condition follows from the first, since the bases of $N^\diamond$ are the
complements of the bases of $N$. Applying ${}^*$ to $B\sqcup([n]\setminus B)^*$ gives
$([n]\setminus B)\sqcup B^*$, which is a basis exactly when $[n]\setminus B$ is a basis of $N$; so
$L_{\mathsf C}$ is self-dual if and only if the complement of every basis of $N$ is a basis of $N$, that
is, $N^\diamond=N$.

(2) The dealer block of $M$ has the columns $(g_d;0)$ and $(0;g^\perp_d)$, which are both nonzero if and
only if $d$ is not a loop of $N$ and not a loop of $N^\diamond$, that is, not a coloop of $N$. A vector $(c\mid c')$ of $W_{\mathsf C}$ is zero at $d$ and $d^*$ exactly when $c_d=c'_d=0$, so $S_d=\{(c\mid c'):c\in\mathrm{sh}_d(\mathsf C),\ c'\in\mathrm{sh}_d(\mathsf C^\perp)\}$, and $\mathrm{sh}_d(\mathsf C^\perp)=\mathrm{pu}_d(\mathsf C)^\perp$.

(3) A vector $(c\mid c')$ of $W_{\mathsf C}$ lies in $E_d\oplus E_A$ if and only if both $c$ and $c'$ are
supported on $A\cup\{d\}$. Hence
$D_d(A)=\{(x\mid z):x\in D^X,\ z\in D^Z\}$, where $D^X=\{c_d:c\in\mathsf C,\ \supp(c)\subseteq A\cup\{d\}\}$
and $D^Z=\{c'_d:c'\in\mathsf C^\perp,\ \supp(c')\subseteq A\cup\{d\}\}$ are subspaces of $\Fq$, so
$\dim_{\Fq}D_d(A)=\dim_{\Fq}D^X+\dim_{\Fq}D^Z$. A word $c'\in\mathsf C^\perp$ is a linear relation
$\sum_ic'_ig_i=0$ among the columns of $G$. So $D^Z\neq0$ if and only if some relation supported on
$A\cup\{d\}$ has $c'_d\neq0$, that is, $g_d\in\Span_{\Fq}\{g_a:a\in A\}$, or $d\in\cl_N(A)$. In the same
way, the words of $\mathsf C=(\mathsf C^\perp)^\perp$ are the relations among the columns of $G^\perp$, and
$D^X\neq0$ if and only if $d\in\cl_{N^\diamond}(A)$.

(4) Now $\cl_N=\cl_{N^\diamond}$, so by (3) $\dim_{\Fq}D_d(A)\in\{0,2\}$; there are no intermediate sets, and
$A$ is authorized if and only if $d\in\cl_N(A)$. In that case there is $c'\in\mathsf C^\perp$ supported on
$A\cup\{d\}$ with $c'_d\neq0$, and $(0\mid c')\in W_{\mathsf C}\cap(E_d\oplus E_A)$ has nonzero dealer part
and admissible support, consisting of starred elements only; so $A\in\Gcirc_{L_{\mathsf C},d}$ by
Proposition~\ref{prop:admissible-lift}. The reverse inclusion is Theorem~\ref{thm:selfdual} below, since
$L_{\mathsf C}$ is self-dual by (1).
\end{IEEEproof}

An explicit CSS encoder is also available. For a surjective dealer, put
$C_0=\mathrm{sh}_d(\mathsf C)$ and $C_1=\mathrm{pu}_d(\mathsf C)$.
Then $C_0\subset C_1$ and $\dim_{\mathbb{F}_{q}} C_1-\dim_{\mathbb{F}_{q}} C_0=1$: the dealer is not a loop, so shortening reduces
dimension by one, and not a coloop, so puncturing preserves it. Choose $g\in C_1\setminus C_0$.
For the lift with positive CSS generators the encoding is
\[ |a\rangle\longmapsto |C_0|^{-1/2}\sum_{c\in C_0}|ag+c\rangle,\qquad a\in\Fq. \]
The cosets are disjoint, so the vectors are orthonormal; translations by $C_0$ and phases from
$C_1^\perp$ fix them. They therefore span the code of Proposition~\ref{prop:css}(2).

Part (4) recovers the result of \cite{SarvepalliRaussendorf2010} that an identically self-dual matroid
gives a perfect scheme whose minimal authorized sets are the sets $C'\setminus\{d\}$ for the circuits
$C'$ of $N$ through $d$.

\subsection{Uniqueness of representability over \(\mathbb{F}_{2}\)}\label{subsec:rigidity}

\begin{theorem}\label{thm:rigidity}
Let $W,W'\le\F_2^{2n}$ be isotropic subspaces of the same dimension $r$ that represent the same symplectic matroid $\cS$. Then $W=W'$.
\end{theorem}
\begin{IEEEproof}
Fix a basis $B$ of $\cS$. Over $\F_2$ the swap of the two coordinates of a pair preserves $\omega$ and
all supports, and transforms representations of a matroid into representations of the relabelled matroid. Applying the swaps of the pairs $i$ with $i^*\in B$, and permuting the pairs, we may assume $B=\{1,\dots,r\}$. Let $H$ be the generator matrix of $W$ with $H[B]=I_r$, and write $H=[X\mid Z]$ with
$X=[\,I_r\ X'\,]$ and $Z=[\,Z_B\ Z'\,]$, where $X',Z'$ contain the columns of the pairs $r+1,\dots,n$ and $Z_B$ the columns $1^*,\dots,r^*$. We show that $\cS$ determines $H$. For $b\in B$ and $\gamma\in J\setminus B$, the set $(B\setminus\{b\})\cup\{\gamma\}$ has determinant $H_{b,\gamma}$. It
is admissible when $\gamma$ belongs to one of the pairs $r+1,\dots,n$, and when $\gamma=b^*$. Hence $\cS$
determines $X'$, $Z'$ and the diagonal of $Z_B$. Let $i\neq j$ in $B$. Isotropy says that $XZ^T=Z_B^T+X'Z'^T$ is symmetric, so the sum $s_{ij}=(Z_B)_{ij}+(Z_B)_{ji}=(X'Z'^T+Z'X'^T)_{ij}$ is determined. The admissible set $(B\setminus\{i,j\})\cup\{i^*,j^*\}$ has determinant $(Z_B)_{ii}(Z_B)_{jj}+(Z_B)_{ij}(Z_B)_{ji}$, so the product $(Z_B)_{ij}(Z_B)_{ji}$ is determined. If $s_{ij}=0$, then $(Z_B)_{ij}=(Z_B)_{ji}$ equals this product. If $s_{ij}=1$, then some pair $k>r$ has
$(X'_{ik},Z'_{ik})\neq(0,0)$; choose $\gamma\in\{k,k^*\}$ with $H_{i,\gamma}=1$. The admissible set $(B\setminus\{i,j\})\cup\{i^*,\gamma\}$ has determinant $(Z_B)_{ii}H_{j,\gamma}+(Z_B)_{ji}$, which determines $(Z_B)_{ji}$, and $(Z_B)_{ij}=(Z_B)_{ji}+1$. Thus $H$, and so $W$, is determined by $\cS$ and
$B$. The same holds for $W'$, so $W=W'$.
\end{IEEEproof}

For Lagrangian matroids this uniqueness is known: Zhukov \cite[Thm.~2.1]{Zhukov2018} gives the
bijection between binary Lagrangian subspaces and binary delta-matroids; see also
\cite{BouchetDuchamp1991}. In a symmetric binary chart, principal minors of orders one and two
determine the matrix. Theorem~\ref{thm:rigidity} supplies a reconstruction proof for arbitrary
isotropic rank, including the nonsymmetric block $Z_B$ when $r<n$.

\begin{example}\label{ex:q3}
Let $q\ge3$, choose $c\in\Fq\setminus\{0,1\}$, and let
$A_c=\left(\begin{smallmatrix}1&1\\1&c\end{smallmatrix}\right)$. All four transversals are bases for both $[I_2\mid I_2]$ and $[I_2\mid A_c]$, since the principal minors $1,1,c,c-1$ of $A_c$ are nonzero; so both represent the Lagrangian matroid $U^*_{2,2}$ in which every transversal is a basis. However
$\rho(\{1\})=1+\cutrk(\{1\})$ equals $1$ for $I_2$ and $2$ for $A_c$, and the stabilizer weight enumerators are $(1+(q-1)z)^2$ and $1+(q^2-1)z^2$. For $[I_2\mid I_2]$ no dealer is surjective. For $[I_2\mid A_c]$ both dealers are surjective, and the other player alone is authorized, by Corollary~\ref{cor:cutrank}; so each dealer gives a perfect scheme. Yet $U^*_{2,2}$ has no circuits, so $\Gcirc_{U^*_{2,2},d}=\emptyset$ for both dealers.

Operational access can vary even with a common surjective dealer. The symmetric matrices
\[
A_{12}=\begin{pmatrix}1&1&0\\1&c&0\\0&0&1\end{pmatrix},\qquad
A_{13}=\begin{pmatrix}1&0&1\\0&1&0\\1&0&c\end{pmatrix}
\]
have every principal minor nonzero, so both $[I_3\mid A_{12}]$ and $[I_3\mid A_{13}]$ represent
$U^*_{3,3}$. Dealer $1$ is surjective in both. The first graph state is an entangled pair on
$\{1,2\}$ tensored with a one-qudit state, while the second has the entangled pair on $\{1,3\}$.
Equivalently, Corollary~\ref{cor:cutrank} gives
\begin{align*}
 \Grec_1(W_{A_{12}})&=\{\{2\},\{2,3\}\},\\
 \Grec_1(W_{A_{13}})&=\{\{3\},\{2,3\}\}.
\end{align*}
Both schemes are perfect, but their labelled access structures differ.
\end{example}

\begin{corollary}\label{cor:dichotomy}
For a prime power $q$ the following are equivalent:
\begin{enumerate}
\item every $\Fq$-representable symplectic matroid has a unique representation up to local Clifford
operations;
\item the pair-rank function of every $\Fq$-linear stabilizer code is determined by its symplectic
matroid;
\item the stabilizer weight enumerator of every $\Fq$-linear stabilizer code is determined by its
symplectic matroid;
\item $q=2$.
\end{enumerate}
\end{corollary}
\begin{IEEEproof}
(4) implies (1), (2) and (3) by Theorem~\ref{thm:rigidity}. If $q\ge3$, Example~\ref{ex:q3} violates
(2) and (3), and also (1), because local Clifford operations preserve supports and hence weight
enumerators.
\end{IEEEproof}

Example~\ref{ex:q3} also shows that a surjective dealer need not lie on a circuit when $q\ge3$. Over
$\F_2$ it must.

\begin{proposition}[Dealers on circuits]\label{prop:nonvacuity}
Let $L$ be a Lagrangian matroid and $d\in[n]$.
\begin{enumerate}
\item No circuit of $L$ meets $J_d$ if and only if $B\triangle J_d$ is a basis for every basis $B$; that
is, $L$ is the direct sum of $L/d$ on $J_{P_d}$ and the matroid on $J_d$ in which both $\{d\}$ and
$\{d^*\}$ are bases.
\item If $W\le\F_2^{2n}$ represents $L$ and $d$ is surjective, then some circuit meets $J_d$; in
particular $\Gcirc_{L,d}\neq\emptyset$.
\end{enumerate}
\end{proposition}
\begin{IEEEproof}
(1) Suppose that no circuit meets $J_d$, let $B$ be a basis and $x$ its element in $J_d$. If
$B'=(B\setminus\{x\})\cup\{x^*\}$ were dependent, it would contain a circuit, which would contain $x^*$
because $B\setminus\{x\}$ is independent. Conversely, suppose that $B\triangle J_d$ is a basis
whenever $B$ is, and let $C$ be a circuit with $x\in C\cap J_d$. Extend the independent set
$C\setminus\{x\}$ to a basis $B$. Either $B$ or $B\triangle J_d$ contains $x$, and hence contains $C$, a
contradiction. The description as a direct sum is a restatement.

(2) Suppose that no circuit meets $J_d$. Then $\{d\}$ is not a circuit, so the column of $d$ is
nonzero and the contraction $L/d$ is
represented by a Lagrangian $W'\le\F_2^{2(n-1)}$ \cite[Lemma~3.3]{GangulyPaperI}. By (1), $L$ is
represented by $W'\oplus\F_2(1\mid1)$, with the second summand placed on the coordinates of $d$; for
this subspace $\rk M_d=1$. By Theorem~\ref{thm:rigidity}, $W$ equals this subspace, so $d$ is not
surjective.
\end{IEEEproof}

\subsection{Circuit-induced versus operational access structures}\label{subsec:comparison}

By Proposition~\ref{prop:admissible-lift} and Theorem~\ref{thm:operational}, the two access structures
are defined by the same vectors, those of $W\cap(E_d\oplus E_A)$ with nonzero dealer part, but in different ways. A set $A$ lies in $\Gcirc_{L,d}$ if one of these vectors has admissible support, and in $\Grec_d(W)$ if their dealer parts span $E_d$. Neither condition implies the other in general.

\begin{theorem}[Self-dual inclusion]\label{thm:selfdual}
Let $L$ be self-dual and represented by a Lagrangian $W$. If a circuit of $L$ meets $J_d$, then $d$ is surjective, and $D_d(B_d(C))=E_d$ for every $C\in\cC_d(L)$. Hence $\Gcirc_{L,d}\subseteq\Grec_d(W)$.
\end{theorem}
\begin{IEEEproof}
Since ${}^*$ maps $\cB(L)$ onto itself, it maps circuits to circuits; so $C^*$ is a circuit, and $B_d(C^*)=B_d(C)$. We may therefore assume $d\in C$. By Lemma~\ref{lem:circuit-vector}, applied to $C$ and to $C^*$, there are $w,w'\in W$ with $\supp_J(w)=C^*$ and $\supp_J(w')=C$. Both lie in $E_d\oplus E_{B_d(C)}$, and their dealer parts are nonzero multiples of $\e_{d^*}$ and $\e_d$. Hence $\pi_d$ is onto and $D_d(B_d(C))=E_d$. By Theorem~\ref{thm:operational}(3), $B_d(C)$ is authorized, and so is every superset of it.
\end{IEEEproof}

In the next theorem we show exactly what is needed to be a perfect quantum secret-sharing scheme beyond Theorem~\ref{thm:selfdual}.

\begin{theorem}\label{thm:comparison}
Let $W$ represent $L$, and let $d$ be surjective.
\begin{enumerate}
\item $\Gcirc_{L,d}\subseteq\Grec_d(W)$ if and only if $D_d(B)=E_d$ for every $B\in\Gmin_{L,d}$.
\item $\Grec_d(W)\subseteq\Gcirc_{L,d}$ if and only if every $A$ with $D_d(A)=E_d$ carries a vector of
$W\cap(E_d\oplus E_A)$ with nonzero dealer part and admissible support.
\item $V_d$ is a perfect scheme with access structure $\Gcirc_{L,d}$ if and only if
$\Gcirc_{L,d}=\Grec_d(W)$ and $\Gint_d(W)=\emptyset$. If $\Gcirc_{L,d}$ is self-dual as an access
structure, that is, $A\in\Gcirc_{L,d}$ if and only if $P_d\setminus A\notin\Gcirc_{L,d}$, then
$\Gcirc_{L,d}=\Grec_d(W)$ already implies $\Gint_d(W)=\emptyset$.
\end{enumerate}
If $L$ is self-dual and $\cC_d(L)\neq\emptyset$, the condition in (1) always holds.
\end{theorem}
\begin{IEEEproof}
(1) follows from the monotonicity of $D_d$ and Theorem~\ref{thm:operational}(3), and (2) from
Proposition~\ref{prop:admissible-lift}. For (3), the first statement is Theorem~\ref{thm:operational}(3).
Assume that $\Gcirc_{L,d}$ is self-dual and equal to $\Grec_d(W)$, and let $A$ be a set that is not
authorized. Then $P_d\setminus A\in\Gcirc_{L,d}=\Grec_d(W)$, so $A$ is forbidden by
Theorem~\ref{thm:operational}. The last sentence is Theorem~\ref{thm:selfdual}.
\end{IEEEproof}

Without self-duality the inclusion can fail, even when $\Gcirc$ is a quantum access structure.

\begin{example}\label{ex:looped}
Over $\F_2$ let $A=\left(\begin{smallmatrix}0&1&0\\1&0&1\\0&1&1\end{smallmatrix}\right)$, the
adjacency matrix of the path $1\,2\,3$ with a loop at $3$, let $L$ be the Lagrangian matroid of
$[I_3\mid A]$, and let $d=1$. The columns $1^*$ and $2$ are equal, so $\{1^*,2\}$ is a circuit and
$\{2\}\in\Gcirc_{L,1}$. Every circuit meeting $J_1$ has $2\in B_1(C)$, so $\Gcirc_{L,1}=\{\{2\},\{2,3\}\}$
is a quantum access structure. But $W\cap(E_1\oplus E_{\{2\}})$ is spanned by the first row
$(1,0,0\mid0,1,0)$, whose support $\{1,2^*\}$ is admissible and whose dealer part spans only a line.
So player $2$ alone is intermediate, and $\Grec_1(W)=\{\{2,3\}\}$. By Theorem~\ref{thm:rigidity}, no
binary representation of $L$ realizes $\Gcirc_{L,1}$ by dealer shortening.
\end{example}

Over $\F_2$ the self-dual graph states have a simple description.

\begin{proposition}\label{prop:graph-selfdual}
Let $A$ be a symmetric matrix over $\Fq$, $W_A=\row[I_n\mid A]$ and $L_A$ its Lagrangian matroid.
\begin{enumerate}
\item The bases of $L_A$ are the transversals $X\sqcup Y^*$, $X\sqcup Y=[n]$, with $\det A[Y,Y]\neq0$, where $\det A[\emptyset,\emptyset]=1$.
\item If $A^2=I$, then $L_A$ is self-dual.
\item If $q=2$, then $L_A$ is self-dual if and only if $A^2=I$. For a simple graph this means that every
vertex has odd degree and any two distinct vertices have an even number of common neighbours.
\item For a simple graph $F$, let $F[K_2]$ be the graph with vertex set $V(F)\times\{1,2\}$ in which
$(v,i)$ and $(w,j)$ are adjacent if $vw$ is an edge of $F$, or if $v=w$ and $i\neq j$; thus each vertex of
$F$ is replaced by two adjacent twins. Its adjacency matrix satisfies $A^2=I$ over $\F_2$. Hence, its binary graph state has a self-dual, secret-sharing
Lagrangian matroid, and $\Gcirc\subseteq\Grec$ for every dealer.
\end{enumerate}
\end{proposition}
\begin{IEEEproof}
(1) Expand the determinant along the unit columns $\e_x$, $x\in X$; what remains is the principal
submatrix $A[Y,Y]$. (2) Let $\mathrm{sw}:\Fq^{2n}\to\Fq^{2n}$, $(a\mid b)\mapsto(b\mid a)$, swap the two
coordinates of every pair. It satisfies $\omega(\mathrm{sw}\,u,\mathrm{sw}\,v)=-\omega(u,v)$, so
$\mathrm{sw}(W_A)=\row[A\mid I_n]$ is Lagrangian. Its column $j$ is the column $j^*$ of $[I_n\mid A]$ and
vice versa, so a transversal $T$ is a basis of $\mathrm{sw}(W_A)$ exactly when $T^*$ is a basis of $L_A$;
that is, $\mathrm{sw}(W_A)$ represents the dual $L_A^*$. If $A^2=I$, then
$\row[A\mid I_n]=\row(A[A\mid I_n])=\row[I_n\mid A]$. (3) If $L_A$ is self-dual, $\mathrm{sw}(W_A)$ and $W_A$
are binary representations of the same matroid, so they are equal by Theorem~\ref{thm:rigidity}. Then
$A$ is invertible and $[A\mid I_n]=A[I_n\mid A]$, so $A^2=I$. The entries of $A^2$ count common
neighbours modulo $2$, and the diagonal entries are the degrees. (4) With
$\mathbf J=\left(\begin{smallmatrix}1&1\\1&1\end{smallmatrix}\right)$, the adjacency matrix of $F[K_2]$ is
$A_F\otimes\mathbf J+I\otimes(\mathbf J+I)$ over $\F_2$, where $A_F$ is that of $F$. Since
$\mathbf J^2=0$, $\mathbf J(\mathbf J+I)=(\mathbf J+I)\mathbf J=\mathbf J$ and $(\mathbf J+I)^2=I$ over
$\F_2$, its square is $A_F^2\otimes0+2(A_F\otimes\mathbf J)+I\otimes I=I$. The remaining claims follow from
Theorems~\ref{thm:selfdual-ss} and~\ref{thm:selfdual}, and the fact that every dealer has a neighbour
in $F[K_2]$, hence lies on the circuit $N(d')\sqcup\{d'^*\}$ of its twin $d'$.
\end{IEEEproof}

Even for self-dual matroids the inclusion of Theorem~\ref{thm:selfdual} can be strict. The graph
$P_3[K_2]$, where $P_3$ is the path on three vertices, gives an example. It is minimal among simple
graph-state instances whose circuit-induced family at the chosen dealer is a quantum access
structure, as verified by Table~\ref{tab:census}.

\begin{theorem}\label{thm:Gstar}
Let $\Gamma^\star$ be the graph on $\{0,\dots,5\}$ with edges $01$, $02$, $05$, $12$, $13$, $14$, $15$,
$25$, $34$, $35$, $45$; it is isomorphic to $P_3[K_2]$, the twin pairs being $\{0,2\}$, $\{1,5\}$,
$\{3,4\}$. Let $L^\star$ be the Lagrangian matroid of its binary graph state, with its unique binary
representation $W$. Then:
\begin{enumerate}
\item $L^\star$ has $20$ circuits, listed in the Appendix; it is self-dual and secret-sharing.
\item For every dealer $d$, $\Gcirc_{L^\star,d}\subsetneq\Grec_d(W)$, and there are exactly $10$
intermediate sets.
\item Let $\iota=(0\,2)(1\,5)(3\,4)$ be the involution exchanging twins. For every $d$ the minimal authorized
sets are the six sets $\{\iota(d)\}\cup T$ with $T$ a $2$-subset of $P_d\setminus\{\iota(d)\}$, and only two
of them are minimal circuit-induced sets. For $d=0$ these are $\{1,2,5\}$ and $\{2,3,4\}$, and
$\{1,2,3\}$ is authorized but contains no circuit-induced set.
\item For every dealer, $Q_d(W)$ is a $[[5,1,1]]_2$ code: the twin $\iota(d)$ alone is intermediate. For
$d=0$, Construction~\ref{con:dealer} followed by row reduction gives, on the players $1,\dots,5$, the
stabilizer matrix
\[ H_0=\left[\begin{array}{ccccc|ccccc}
1&0&0&0&1&1&0&0&0&1\\
0&1&0&0&1&0&1&1&1&1\\
0&0&1&0&0&1&0&0&1&1\\
0&0&0&1&0&1&0&1&0&1
\end{array}\right]. \]
\end{enumerate}
Consequently, for no dealer is $\Gcirc_{L^\star,d}$ the access structure of the dealer-shortened code of a
binary representation of $L^\star$.
\end{theorem}
\begin{IEEEproof}
Self-duality and the secret-sharing property follow from Proposition~\ref{prop:graph-selfdual}(4). The
remaining statements are finite verifications, described in the Appendix: the circuits are obtained
from Theorem~\ref{thm:graph-circuits}, and the families $\Grec_d$ and $\Gint_d$ from
Corollary~\ref{cor:cutrank}. For example, for $d=0$, $\cutrk(\{1,2,3\})=3$ and $\cutrk(\{0,1,2,3\})=2$,
so $\{1,2,3\}$ is authorized, while $\cutrk(\{1,3,4\})=\cutrk(\{0,1,3,4\})=2$, so $\{1,3,4\}$ is
intermediate. For (4), the rows of the twins $d$ and $\iota(d)$ agree outside $\{d,\iota(d)\}$, so
$\cutrk(\{\iota(d)\})=\cutrk(\{d,\iota(d)\})=1$ and $\dim_{\Fq} D_d(\{\iota(d)\})=1$. The final statement follows
from Theorem~\ref{thm:rigidity}.
\end{IEEEproof}

\begin{table}[t]
\centering
\caption{Binary graph states of connected simple graphs on $n$ vertices: the pairs $(\Gamma,d)$ for
which $\Gcirc$ is a quantum access structure (q.a.s.), the number of these with $\Gcirc\subseteq\Grec$
and with $\Gcirc=\Grec$, and the number of secret-sharing $L(\Gamma)$ (in parentheses: those with
$\Gcirc=\Grec$ for every dealer).}\label{tab:census}
\begin{tabular}{@{}ccccccc@{}}
\toprule
$n$ & graphs & pairs $(\Gamma,d)$ & q.a.s. & $\subseteq$ & $=$ & secret-sharing\\
\midrule
2 & 1 & 2 & 2 & 2 & 2 & 1 (1)\\
3 & 2 & 6 & 3 & 3 & 3 & 1 (1)\\
4 & 6 & 24 & 8 & 8 & 8 & 1 (1)\\
5 & 21 & 105 & 29 & 29 & 29 & 1 (1)\\
6 & 112 & 672 & 156 & 156 & 120 & 4 (3)\\
\bottomrule
\end{tabular}
\end{table}

Table~\ref{tab:census} shows that for graph states of simple graphs, six qubits is the smallest size at
which a circuit-induced quantum access structure differs from the operational one. The one-vertex
graph does not qualify: its dealer circuit has empty player support. For a
disconnected graph, both structures, and the condition of Proposition~\ref{prop:nocloning}, are
determined by the component of the dealer, because circuits and cut-ranks split over components. For
simple graphs on at most six vertices, inclusion always holds when $\Gcirc$ is a quantum access
structure, in contrast with the looped graph of Example~\ref{ex:looped}. For Sarvepalli's six-qubit
example \cite[Sec.~IV]{Sarvepalli2011} the two structures coincide and form the $((3,5))$ threshold
structure, realized by a $[[5,1,3]]_2$ dealer-shortened code.

\begin{remark}[Answers to Questions~3 and~4]\label{rem:Q34}
Question~3 is answered by Proposition~\ref{prop:nocloning} and Corollary~\ref{cor:secret-sharing}, with
the sufficient condition of Theorem~\ref{thm:selfdual-ss} for self-dual matroids, by
Proposition~\ref{prop:admissible-lift} for represented matroids, and in graph language by
Theorem~\ref{thm:graph-circuits} and Corollary~\ref{cor:girth5}. For Question~4,
the code attached to a represented Lagrangian matroid and a surjective dealer is $Q_d(W)$, given
explicitly by Theorem~\ref{thm:Hd} and Construction~\ref{con:dealer}, with encoding and recovery as in
Theorem~\ref{thm:encoding}, and implemented by the stabilizer state (Proposition~\ref{prop:choi}). Its
access structure $\Grec_d(W)$ is computed exactly by Theorem~\ref{thm:operational}. It contains
$\Gcirc_{L,d}$ if $L$ is self-dual (Theorem~\ref{thm:selfdual}) but not in general
(Example~\ref{ex:looped}); it can be strictly larger even for self-dual secret-sharing binary $L$
(Theorem~\ref{thm:Gstar}); and it is determined by $L$ for every $L$ if and only if $q=2$
(Theorem~\ref{thm:rigidity} and Example~\ref{ex:q3}). Thus the precise object behind the code is the
represented pair $(L,W)$, not the matroid $L$ alone, except over $\F_2$.
For self-dual $L$, Corollary~\ref{cor:hybrid-circuit} realizes $\Gcirc_{L,d}$ perfectly after
adding classical shares. A pure dealer-shortened realization has exactly the conditions in
Theorem~\ref{thm:comparison}; the hybrid result does not remove those restrictions.
\end{remark}

\section{The symplectic support polynomial}\label{sec:enum}

For an isotropic $S=\row(M)$ of dimension $r$, let $W_S(z)=\sum_{s\in S}z^{\wt(s)}$ and
$W_{S^\perp}(z)=\sum_{u\in S^\perp}z^{\wt(u)}$ be the stabilizer and normalizer weight enumerators.

\begin{definition}\label{def:support-poly}
The \emph{symplectic support polynomial} of $M$ is
\[ \cP_M(\alpha,\beta,\gamma)=\sum_{A\subseteq[n]}\alpha^{|A|}\beta^{\rho_M(A)}\gamma^{\rho_M(\bar A)}. \]
\end{definition}

It depends only on $S$ and aggregates the triples
$(|A|,\rho_M(A),\rho_M(\bar A))$. It therefore determines the enumerators below, but does not retain
the identity of each set $A$. The labelled refinement
\[ \widetilde{\cP}_M(\mathbf t,\beta,\gamma)=
\sum_{A\subseteq[n]}\Big(\prod_{i\in A}t_i\Big)\beta^{\rho_M(A)}\gamma^{\rho_M(\bar A)} \]
does retain the complete pair-rank function. For example, interchanging qudits in the spaces
$\Span_{\F_2}\{(1,0\mid0,0)\}$ and $\Span_{\F_2}\{(0,1\mid0,0)\}$ leaves $\cP_M$ equal to
$\gamma+\alpha\beta+\alpha\gamma+\alpha^2\beta$, while their stabilizers supported on qudit $1$
have different dimensions.

\begin{theorem}\label{thm:enumerator}
\begin{align*}
W_S(z)&=q^{r}(1-z)^n\,\cP_M\Big(\frac{z}{1-z},\,1,\,q^{-1}\Big),\\
W_{S^\perp}(z)&=(1-z)^n\,\cP_M\Big(\frac{q^2z}{1-z},\,q^{-1},\,1\Big).
\end{align*}
When $r<n$, $\cP_M$ determines the distance,
$d(S)=\min\{j\ge1:[z^j]\big(W_{S^\perp}-W_S\big)>0\}$, and
$W_{S^\perp}(z)=q^{-r}\big(1+(q^2-1)z\big)^nW_S\big(\frac{1-z}{1+(q^2-1)z}\big)$.

Here we write $[z^j]F(z)$ for the coefficient of $z^j$ in $F(z)$;
\[
[z^j]F(z):=
\begin{cases}
a_j, & 0\le j\le N,\\
0,   & j>N,
\end{cases}
\]
where $F(z)=\sum_{\ell=0}^{N}a_\ell z^\ell$ is a polynomial and $j\ge0$ is an integer.
In particular, since $S\subseteq S^\perp$, the quantity
\[
[z^j]\bigl(W_{S^\perp}(z)-W_S(z)\bigr)
\]
counts the vectors in $S^\perp\setminus S$ having symplectic
weight $j$.
\end{theorem}
\begin{IEEEproof}
For a subspace $U\le\Fq^{2n}$ and $v\in U$, the identity
$z^{\wt(v)}=\sum_{A\supseteq\supp(v)}z^{|A|}(1-z)^{n-|A|}$ gives
$\sum_{v\in U}z^{\wt(v)}=\sum_{A\subseteq[n]}|U\cap E_A|\,z^{|A|}(1-z)^{n-|A|}$. Insert
$|S\cap E_A|=q^{r-\rho_M(\bar A)}$ and $|S^\perp\cap E_A|=q^{2|A|-\rho_M(A)}$ from
Lemma~\ref{lem:counting}. The identity between $W_S$ and $W_{S^\perp}$ is the MacWilliams transform for
additive codes over an alphabet of size $q^2$, with the duality given by $\tr_{\mathbb{F}_{q}/\mathbb{F}_{p}}\circ\omega$
\cite{ShorLaflamme1997,Rains1999,KetkarEtAl2006}.
\end{IEEEproof}

For the theta graph of Example~\ref{ex:theta}, the enumerators displayed there agree with the
theorem, and the first index at which they differ is $3$, the distance.

\begin{remark}
The function $\rho_M$ is the rank function of the $2$-polymatroid formed by the column pairs of $M$, and
Theorem~\ref{thm:enumerator} is Greene's argument \cite{Greene1976}, M\"obius inversion over the subcodes
supported on the sets $A$, applied to this $2$-polymatroid viewed as an additive code over an alphabet
of size $q^2$. Tutte invariants of $2$-polymatroids are studied in \cite{OxleyWhittle1993}, and
Greene-type theorems for polymatroids and group codes in \cite{SwartzWentworthNiceXue2024}. What is
specific here is the relation with the symplectic matroid, described next.
\end{remark}

\begin{corollary}\label{cor:enum-invariance}
\begin{enumerate}
\item $\cP_M$ is invariant under $\Fq$-linear local Clifford operations, and $W_S$, $W_{S^\perp}$ under all
local Clifford operations. For a graph state $W_A=\row[I_n\mid A]$, put $h(Y)=\cutrk_A(Y)$; then
\[ \cP(\alpha,\beta,\gamma)=\sum_{Y\subseteq[n]}\alpha^{|Y|}\beta^{|Y|+h(Y)}
\gamma^{\,n-|Y|+h(Y)}. \]
\item If $q=2$, then $\cP_M$, $W_S$, $W_{S^\perp}$, and the enumerators of all dealer-shortened codes of a
Lagrangian $W$ depend only on the symplectic matroid of $M$.
\item If $q\ge3$, no function of the symplectic matroid alone determines $W_S$ (Example~\ref{ex:q3}).
\end{enumerate}
\end{corollary}
\begin{IEEEproof}
(1) Pair ranks are invariant under invertible transformations of the column pairs, and local
Clifford operations preserve supports; for graph states insert Lemma~\ref{lem:cutrank} and
$\cutrk_A(\bar Y)=\cutrk_A(Y)$. (2) follows from Theorem~\ref{thm:rigidity}, and (3) from
Example~\ref{ex:q3}.
\end{IEEEproof}

\begin{remark}[Answer to Question~5 and relation to the paper \cite{GangulyPaperI}]\label{rem:Q5}
For a fixed representation, $\cP_M$ answers the enumerator question; for $q\ge3$, Corollary~\ref{cor:enum-invariance}(3) rules out such an answer from the matroid alone. The paper \cite{GangulyPaperI} studies the
admissible corank--nullity polynomial $T_{\cS}(u,v)=\sum_Xu^{r-r_{\cS}(X)}v^{|X|-r_{\cS}(X)}$, summed over
the admissible sets $X$, which is an invariant of every symplectic matroid. It uses only ranks of
admissible sets, whereas $\cP_M$ uses the ranks of the sets $J_A$, which contain both elements of every
pair they meet. By Corollary~\ref{cor:dichotomy} these ranks are determined by the matroid exactly when
$q=2$. The two polynomials are therefore complementary: $T_{\cS}$ is intrinsic to the matroid, and
$\cP_M$ is intrinsic to the code.
\end{remark}

\section{Conclusion and open problems}\label{sec:concl}

The symplectic matroid determines admissible column ranks, while pair ranks govern supported logical operators and recovery. Over $\F_2$, uniqueness of representability makes all pair ranks recoverable from the labelled matroid; over larger finite fields this implication fails. Its converse already fails
over $\F_2$: the path $P_3$ and triangle $K_3$ have cut rank one on every nonempty proper subset, and hence identical pair ranks, but different admissible bases. Thus pair ranks do not determine the symplectic matroid.

We prove the $3$-connected simple-graph case of the graphical parameter question, give exact criteria for circuit and operational access, construct the canonical dealer-shortened code, and identify when that code realizes the circuit-induced structure. A perfect hybrid realization is available under the inclusion hypothesis of Corollary~\ref{cor:hybrid-circuit}; a pure stabilizer realization requires the additional conditions of Theorem~\ref{thm:comparison}. The support polynomial yields both weight enumerators, with a matroid-only determination in the binary case. Appendix~\ref{app:envelopes} preserves the ordinary-envelope viewpoint while distinguishing ordinary rank operations from physical stabilizer constructions. The following problems remain.

\begin{enumerate}
\item Is the distance unbounded over the representations of $\GraphS{G}$ when $G$ ranges over the
subdivisions of $3$-connected graphs? Which rates and distances can graphical codes achieve in
characteristic~$2$?
\item For simple graphs $\Gamma$, does the condition that $\Gcirc_{L(\Gamma),d}$ is a quantum access
structure imply $\Gcirc_{L(\Gamma),d}\subseteq\Grec_d$? This holds for at most six vertices
(Table~\ref{tab:census}) and fails for graphs with loops (Example~\ref{ex:looped}).
\item Characterize the binary graph states with $\Gcirc_{L,d}=\Grec_d$ for every dealer, for example in
terms of local complementation and the condition $A^2=I$ of Proposition~\ref{prop:graph-selfdual}.
\item For $q\ge3$, which representations of a Lagrangian matroid maximize the operational access
structure, or make the scheme perfect?
\end{enumerate}

\appendices
\section{Computations}\label{app:computations}
Theorem~\ref{thm:Gstar}, the computation Table~\ref{tab:census}, the statements about Sarvepalli's six-qubit example in Section~\ref{subsec:comparison}, and the remark after Lemma~\ref{lem:circuit-bound} are finite verifications. Other new assertions are proved in the text; imported results are cited explicitly. The programs, written in Python~3 with the packages \texttt{galois}, \texttt{numpy}, \texttt{networkx} and \texttt{sympy}, are provided as supplementary material together with their output. A second
program, \texttt{verify\_stdlib.py}, uses only the Python standard library and rechecks Theorem~\ref{thm:Gstar}, Examples~\ref{ex:K4}--\ref{ex:looped}, and Table~\ref{tab:census} for $n\le5$; on the connected simple graphs with at most five vertices it also checks Theorem~\ref{thm:graph-circuits}(1) and Lemma~\ref{lem:orthogonality}.

For Table~\ref{tab:census}, the connected simple graphs on $n\le6$ vertices are taken from the atlas of
graphs, one graph per isomorphism class. For each graph and dealer, the circuits are computed from Theorem~\ref{thm:graph-circuits} and also directly from the minors of $[I_n\mid A]$, which agree; $\Gcirc$ is formed as in Definition~\ref{def:circ}; and $\Grec$ and $\Gint$ are computed from the pair-rank formula of Theorem~\ref{thm:operational}. The same programs verify Theorem~\ref{thm:Gstar}, Example~\ref{ex:looped}, and the statements about Sarvepalli's six-qubit example. For the remark after Lemma~\ref{lem:circuit-bound}, the distance of $Q_d(W)$ was compared with the smallest circuit meeting
$J_d$ for random Lagrangian subspaces over $\F_2,\dots,\F_5$; among the $791$ surjective-dealer instances having a dealer circuit, the bound was attained in $397$ and strict in $394$. The remaining $31$ surjective-dealer instances had no dealer circuit.

The additional script \texttt{verify\_revision.py} checks the strengthened nonbinary example, the information-channel and hybrid calculations on a small encoding, coherent odd-characteristic phases, the explicit encoder below, and the envelope ranks and recurrence in Appendix~\ref{app:envelopes}. The independent standard-library audit \texttt{audit\_core.py} also regenerates the full six-vertex verifications and all $19926$ small binary isotropic spaces. Current run outputs and the execution environment accompany the programs.

The $20$ circuits of $L^\star$ in Theorem~\ref{thm:Gstar} are the sets $\mathcal{O}(Y)\sqcup Y^*$ for the following primitive Eulerian sets $Y$: the six singletons; the four pairs $\{0,3\}$, $\{0,4\}$, $\{2,3\}$, $\{2,4\}$; the eight triples $\{0,1,2\}$, $\{0,1,5\}$, $\{0,2,5\}$, $\{1,2,5\}$, $\{1,3,4\}$, $\{1,3,5\}$, $\{1,4,5\}$, $\{3,4,5\}$; and the two sets $\{0,1,2,3,4\}$ and $\{0,2,3,4,5\}$.

For $d=0$, write $X_i=X_i(1)$, $Z_i=Z_i(1)$ and $Y_i=iX_iZ_i$. An operator-level verification of Theorem~\ref{thm:Gstar}(4) is given by the commuting positive generators
\[
 Y_1Y_5,\quad Y_2Z_3Z_4Y_5,\quad Z_1X_3Z_4Z_5,\quad Z_1Z_3X_4Z_5.
\]
Their Pauli labels are the rows of $H_0$. The unphased
$XZ$ products corresponding to the first two rows of $H_0$
must be multiplied by $-1$ to obtain the displayed stabilizer
generators, since $Y_j=iX_jZ_j$ and each of these two products
contains two factors of the form $X_jZ_j$.
Choose $\overline X=Z_1Z_2Z_5$ and $\overline Z=X_1Z_2Z_3Z_4Z_5$. A normalized encoder is
\[
 |s\rangle\longmapsto 2^{-5/2}\sum_{b\in\F_2^5}
 (-1)^{Q(b)+s(b_1+b_2+b_5)}|b\rangle,\qquad s\in\F_2,
\]
where
\[
 Q(b)=b_1b_2+b_1b_3+b_1b_4+b_1b_5+b_2b_5+b_3b_4+b_3b_5+b_4b_5.
\]
Each generator fixes these states, $\overline X$ interchanges them, and $\overline Z$ has eigenvalue $(-1)^s$. These claims follow by substituting the relevant bit flip into $Q$.
The amplitudes have constant modulus $2^{-5/2}$ and the relative linear character is nontrivial, proving normalization and orthogonality. The operator $Y_2$ commutes with every generator and anticommutes with $\overline X$, hence is a nontrivial logical operator of weight one.

The remaining checks are noted in the supplementary ``Verification manual'. They cover counting and MacWilliams identities, CSS codes, admissible lifts, dealer shortening, supported logical spaces, coherent phases and Choi identities over small finite fields; Hilbert-space checks use $q=2,3$. Uniqueness of \(\mathbb{F}_{2}\)-representability was exhaustively checked on all $19926$ nonzero isotropic subspaces of $\F_2^{2n}$, $n\le4$, in every rank. Further searches on $K_4$, $K_5$, $K_{3,3}$ and $W_5$ over $\F_4,\F_8$ confirm Theorems~\ref{thm:weight-one} and~\ref{thm:barrier}. General results rely on the proofs in the text, rather than on extrapolation from these bounded computations.

\section{Ordinary envelopes, rank operations and admissible polynomials}\label{app:envelopes}
In this section we give the ordinary-envelope constructions from \cite{GangulyPaperI} alongside the operational invariants. Their objects are ordinary matroids with a fixed involution; isotropic representability is an additional condition. Our basis-compatible envelopes below are not assumed to satisfy the stronger flat conditions in Raz's theory of ranked symplectic matroids \cite{Raz2024}.

\subsection{Compatibility and the intrinsic admissible rank}

\begin{definition}\label{def:envelope}
A pair-compatible ordinary envelope of a rank-$r$ symplectic matroid $\cS$ on $J$ is an ordinary
matroid $N$ on $J$ with $r_N(J)=r$ and
$\mathcal B(N)\cap\mathcal A_r(J)=\cB(\cS)$. It is sheltering if
$r_N(X)=r_{\cS}(X)$ for every admissible $X$.
\end{definition}

The vector matroid of any isotropic representation is sheltering by Lemma~\ref{lem:intrinsic}.
For every pair-compatible envelope and admissible $X$, one has
$r_N(X)\ge r_{\cS}(X)$: each $X\cap B$, $B\in\cB(\cS)$, is independent in $N$.
Define
\begin{align*}
 T_N(u,v)&=\sum_{X\in\mathcal A(J)}u^{r_N(J)-r_N(X)}v^{|X|-r_N(X)},\\
 T_{\cS}(u,v)&=\sum_{X\in\mathcal A(J)}u^{r-r_{\cS}(X)}v^{|X|-r_{\cS}(X)}.
\end{align*}
Then $T_N=T_{\cS}$ if and only if $N$ is sheltering. Sufficiency is termwise equality.
For necessity, evaluate at $u=v=2$: each term for $N$ is at most its term for $\cS$, and is
strictly smaller if $r_N(X)>r_{\cS}(X)$. This also explains why basis compatibility alone does
not give the intrinsic polynomial. These intrinsic-rank and sheltering results are developed
further in \cite{GangulyPaperI}.

Subset size can be retained by the graded polynomial
\[T_N(z;u,v)=\sum_{X\in\mathcal A(J)}z^{|X|}u^{r-r_N(X)}v^{|X|-r_N(X)}.\]
Its rank-$r$ layer satisfies
\[
 [z^r]T_N(z;u,v)=\sum_{X\in\mathcal A_r(J)}(uv)^{r-r_N(X)}.
\]
For a sheltering vector envelope, setting $u=x-1$, $v=1$ gives Sarvepalli's restricted
Tutte--Martin polynomial \cite[Sec.~III-E]{Sarvepalli2011}. The restriction to $|X|=r$ is a
coefficient extraction, not in general a specialization of the ungraded $T_N$.
The pair polynomial $\cP_M$ instead uses $r_N(J_A)$ and $r_N(J_{\bar A})$.
Thus ordinary envelopes supply a common language for both polynomials, but the ranks needed
for the coding enumerators are on generally inadmissible unions of pairs.

On any fixed finite paired ground set, every nonempty family of pair-compatible envelopes
has a weak-order minimal member, since there are finitely many ordinary matroids.
Here $N_1\preceq N_2$ means $r_{N_1}(X)\le r_{N_2}(X)$ for all $X\subseteq J$.
This observation proves neither universal existence of an envelope nor uniqueness of a
minimal member, and a minimal member need not be a least member. Accordingly, no canonical
$N_{\min}(\cS)$ is assumed in general. A least envelope does exist in the uniform case below.

\subsection{Higgs lifts, elongation and truncation}

For ordinary matroids $Q,N$ on the same set with $Q$ a quotient of $N$, the $i$th Higgs lift has
rank $\min\{r_N(X),r_Q(X)+i\}$, $0\le i\le r_N(J)-r_Q(J)$
\cite{Higgs1968,Oxley2011}. Here quotient means that every flat of $Q$ is a flat of $N$.
Taking $N$ free gives the $t$-fold elongation of an ordinary matroid $M$:
\begin{gather*}
 \mathcal I(\operatorname{Elong}_tM)=\{I:|I|-r_M(I)\le t\},\\
 r_{\operatorname{Elong}_tM}(X)=\min\{|X|,r_M(X)+t\}.
\end{gather*}
For the special case used here these formulas have a direct proof. Nullity is nondecreasing,
so the displayed independence system is hereditary. If $I,K$ belong to it and $|I|<|K|$,
then either the nullity of $I$ is less than $t$, in which case any new element augments it, or
it is $t$. In the latter case failure of augmentation would put $K\setminus I$ in
$\cl_M(I)$, giving $|K|-r_M(K)>|I|-r_M(I)=t$, a contradiction.
A basis of $M|X$, together with up to $t$ additional elements of $X$, proves the rank formula.

\begin{proposition}[Preservation under truncated elongation]\label{prop:envelope-gap}
Let $M$ be a rank-$r$ pair-compatible envelope of $\cS$, and define
$M^{(0)}=M$ and $M^{(t+1)}=\operatorname{Trunc}_r(\operatorname{Elong}_1M^{(t)})$.
Then
\[
 r_{M^{(t)}}(X)=\min\{r,|X|,r_M(X)+t\}.
\]
It envelopes the same $\cS$ if and only if
$r-r_M(B)>t$ for every $B\in\mathcal A_r(J)\setminus\cB(\cS)$.
\end{proposition}
\begin{IEEEproof}
The rank formula follows by induction, since truncation takes the minimum with $r$.
For $|B|=r$, the set $B$ is a new basis exactly when $r-r_M(B)\le t$.
Old bases remain bases, so the stated condition is precisely the exclusion of new admissible bases.
\end{IEEEproof}

\begin{proposition}[Uniform envelope hierarchy]\label{prop:uniform-envelopes}
For $1\le r\le n$, put $P_n=\bigoplus_{i=1}^nU_{1,\{i,i^*\}}$ and
$P_{n,r}=\operatorname{Trunc}_rP_n$. Write $p(X)$ for the number of pairs met by $X$ and
$c(X)$ for the number contained in $X$, so $|X|=p(X)+c(X)$.
For $0\le t\le\lfloor r/2\rfloor$, the truncated elongations $M_{n,r}^{(t)}$ of $P_{n,r}$ have
\begin{gather*}
 r_t(X)=\min\{r,|X|,p(X)+t\},\\
 \mathcal I(M_{n,r}^{(t)})=\{I:|I|\le r,\ c(I)\le t\}.
\end{gather*}
They envelope $U^*_{r,n}$, and form a strict weak-order chain from $P_{n,r}$ to the ordinary
uniform matroid $U_{r,J}$, except when $r=1$, when there is only one term.
Moreover $P_{n,r}$ is the least pair-compatible envelope of $U^*_{r,n}$.
\end{proposition}
\begin{IEEEproof}
The rank formula is Proposition~\ref{prop:envelope-gap}, since $r_{P_{n,r}}(X)=\min(r,p(X))$.
The independence formula follows using $|X|=p(X)+c(X)$. Admissible $r$-sets have $c=0$ and
are therefore bases at every stage. The union of $t+1$ complete pairs has rank $2t+1$ at stage
$t$ and $2t+2$ at stage $t+1$ whenever $t<\lfloor r/2\rfloor$, proving strictness.
Every set of size at most $r$ contains at most $\lfloor r/2\rfloor$ complete pairs, proving
the final uniform term. For any other envelope $N$, choose an admissible subset of $X$ of size
$\min(r,p(X))$ and extend it to an admissible $r$-set. It is a subset of a basis of $N$, so
$r_N(X)\ge\min(r,p(X))=r_{P_{n,r}}(X)$.
\end{IEEEproof}

The least envelope has the explicit closure
\[
 \cl_{P_{n,r}}(X)=
 \begin{cases}J_{\varphi(X)},&p(X)<r,\\J,&p(X)\ge r.\end{cases}
\]
Indeed, below rank $r$ an element raises rank exactly when it belongs to a new pair.
At $t=1$ the rank is $\min\{r,p(X)+[c(X)>0]\}$, so this first elongation is already uniform
for $r\le3$. For $n=r=4$, the ranks of $J_{\{1,2\}}$ along the hierarchy are $2,3,4$, whereas
an admissible transversal has rank four throughout.

For all members of the hierarchy, $r_t(X)=\min(r,|X|)$ on admissible sets. Consequently
\[
 T_{M_{n,r}^{(t)}}(u,v)=
 \sum_{s=0}^{r}2^s\binom ns u^{r-s}
 +\sum_{s=r+1}^{n}2^s\binom ns v^{s-r},
\]
independently of $t$. In particular, $T_N(1,1)=3^n$ for every paired ordinary matroid on $n$
pairs. The values $T_N(1,0)$, $T_N(0,1)$ and $T_N(0,0)$ count, respectively, admissible
independent sets, spanning sets and bases. For a direct sum on disjoint paired sets,
$T_{N_1\oplus N_2}=T_{N_1}T_{N_2}$, since both size and rank add.
The uniform hierarchy is defined combinatorially; neither its ordinary representability nor
its isotropic representability follows from the construction. This distinction also applies
to Higgs lifts of other envelopes.

\subsection{Relaxation and a three-branch recurrence}

If an inadmissible $H$ is a circuit-hyperplane of a rank-$r$ envelope $M$, relaxing $H$ adds
exactly $H$ to the basis family \cite{Oxley2011}. To recall why this is a matroid, an exchange
from $H$ to a basis $B$ works because $H\setminus\{e\}$ is independent and every element
outside $H$ extends it to a basis. For an exchange from $B$ to $H$, the hyperplane
$\cl_M(B\setminus\{e\})$ either equals $H$, yielding the new basis $H$, or omits an element
of $H$, yielding an old basis. Its only changed rank is $r(H):r-1\mapsto r$.
Therefore relaxation preserves the admissible bases, all admissible ranks, and $T_M$.
This can be iterated whenever another inadmissible circuit-hyperplane is available; it need
not generate all envelopes or preserve representability.

For example, relax $\{1,1^*\}$ in $P_{2,2}$. Its pair ranks become $2$ on $\{1\}$ and $1$ on
$\{2\}$. It still envelopes $U^*_{2,2}$ and has an ordinary binary representation with columns
$\e_1,\e_2,\e_1+\e_2,\e_2$. It has no isotropic representation over any field: a Lagrangian
two-qudit representation must satisfy $\rho(\{1\})=\rho(\{2\})$ by Lemma~\ref{lem:counting}.
Thus even ordinary representability of a modified envelope does not produce a stabilizer code.

Finally, the state sum itself has an exact recurrence, without any compatibility assumption on
the minors. For an ordinary $N$ on $J$, set $r=r_N(J)$, fix $e=\{a,a^*\}$, and put
$N_0=N\setminus e$, $N_a=(N/a)\setminus a^*$ and $N_{a^*}=(N/a^*)\setminus a$.
All three have paired ground set $J\setminus e$. Write $r_b=r(N_b)$ and
$\varepsilon_b=r_N(\{b\})$, $b=a,a^*$. Then
\begin{equation}\label{eq:envelope-recursion}
 T_N=u^{r-r_0}T_{N_0}
 +\sum_{b\in\{a,a^*\}}u^{r-r_b-\varepsilon_b}v^{1-\varepsilon_b}T_{N_b}.
\end{equation}
Indeed, admissible sets meet $e$ in $\emptyset$, $\{a\}$ or $\{a^*\}$.
For $X\subseteq J\setminus e$, contraction gives
$r_N(X\cup\{b\})=\varepsilon_b+r_{N_b}(X)$, proving each summand. All displayed exponents
are nonnegative. If both elements are loops, all three minors coincide and the formula is
$(1+2v)T_{N_0}$. The minors are ordinary paired matroids; an interpretation as symplectic
envelopes requires checking their admissible bases. This mate-deleted contraction is distinct
from contractions that additionally excise flats.

For $P_{n,r}$ with $r\ge1$, \eqref{eq:envelope-recursion} reduces to
\[
 T_{P_{n,r}}=u^{\delta_{r,n}}T_{P_{n-1,\min(r,n-1)}}+2T_{P_{n-1,r-1}},
\]
where $\delta_{r,n}$ is one if $r=n$ and zero otherwise. Here $P_{n,0}$ denotes the rank-zero matroid
and $P_{0,0}$ the empty matroid; for $r=0$ the loop factor applies.
Already $P_{1,1}$ gives $T=u+2$, demonstrating why an unweighted three-term recurrence
without rank-loss factors would be incorrect.

\bibliographystyle{IEEEtran}
\bibliography{references}

@book{BorovikGelfandWhite2003,
  author    = {Borovik, Alexandre V. and Gelfand, Israel M. and White, Neil},
  title     = {Coxeter Matroids},
  series    = {Progress in Mathematics},
  volume    = {216},
  publisher = {Birkh{\"a}user},
  address   = {Boston},
  year      = {2003}}

@misc{Sarvepalli2011,
  author = {Sarvepalli, Pradeep},
  title  = {Quantum codes and symplectic matroids},
  year   = {2011},
  note   = {arXiv:1104.1171 [quant-ph]}}

@article{SarvepalliRaussendorf2010,
  author  = {Sarvepalli, Pradeep and Raussendorf, Robert},
  title   = {Matroids and quantum-secret-sharing schemes},
  journal = {Phys. Rev. A},
  volume  = {81},
  pages   = {052333},
  year    = {2010}}

@article{Chow2003,
  author  = {Chow, Timothy Y.},
  title   = {Symplectic matroids, independent sets, and signed graphs},
  journal = {Discrete Math.},
  volume  = {263},
  pages   = {35--45},
  year    = {2003}}

@article{CleveGottesmanLo1999,
  author  = {Cleve, Richard and Gottesman, Daniel and Lo, Hoi-Kwong},
  title   = {How to share a quantum secret},
  journal = {Phys. Rev. Lett.},
  volume  = {83},
  pages   = {648--651},
  year    = {1999}}

@article{Gottesman2000,
  author  = {Gottesman, Daniel},
  title   = {Theory of quantum secret sharing},
  journal = {Phys. Rev. A},
  volume  = {61},
  pages   = {042311},
  year    = {2000}}

@misc{Smith2000,
  author = {Smith, Adam D.},
  title  = {Quantum secret sharing for general access structures},
  year   = {2000},
  note   = {arXiv:quant-ph/0001087}}

@article{MarkhamSanders2008,
  author  = {Markham, Damian and Sanders, Barry C.},
  title   = {Graph states for quantum secret sharing},
  journal = {Phys. Rev. A},
  volume  = {78},
  pages   = {042309},
  year    = {2008}}

@article{Sarvepalli2012,
  author  = {Sarvepalli, Pradeep},
  title   = {Nonthreshold quantum secret-sharing schemes in the graph-state formalism},
  journal = {Phys. Rev. A},
  volume  = {86},
  pages   = {042303},
  year    = {2012}}

@inproceedings{JavelleMhallaPerdrix2012,
  author    = {Javelle, J{\'e}r{\^o}me and Mhalla, Mehdi and Perdrix, Simon},
  title     = {New protocols and lower bounds for quantum secret sharing with graph states},
  booktitle = {Theory of Quantum Computation, Communication, and Cryptography (TQC 2012)},
  series    = {Lecture Notes in Comput. Sci.},
  volume    = {7582},
  publisher = {Springer},
  pages     = {1--12},
  year      = {2013}}

@article{MarinMarkham2013,
  author  = {Marin, Anne and Markham, Damian},
  title   = {Equivalence between sharing quantum and classical secrets and error correction},
  journal = {Phys. Rev. A},
  volume  = {88},
  pages   = {042332},
  year    = {2013}}

@article{KetkarEtAl2006,
  author  = {Ketkar, Avanti and Klappenecker, Andreas and Kumar, Santosh and Sarvepalli, Pradeep K.},
  title   = {Nonbinary stabilizer codes over finite fields},
  journal = {IEEE Trans. Inf. Theory},
  volume  = {52},
  number  = {11},
  pages   = {4892--4914},
  year    = {2006}}

@article{CRSS1998,
  author  = {Calderbank, A. Robert and Rains, Eric M. and Shor, Peter W. and Sloane, Neil J. A.},
  title   = {Quantum error correction via codes over {GF}(4)},
  journal = {IEEE Trans. Inf. Theory},
  volume  = {44},
  number  = {4},
  pages   = {1369--1387},
  year    = {1998}}

@article{ShorLaflamme1997,
  author  = {Shor, Peter and Laflamme, Raymond},
  title   = {Quantum analog of the {M}ac{W}illiams identities for classical coding theory},
  journal = {Phys. Rev. Lett.},
  volume  = {78},
  pages   = {1600--1602},
  year    = {1997}}

@article{Rains1999,
  author  = {Rains, Eric M.},
  title   = {Nonbinary quantum codes},
  journal = {IEEE Trans. Inf. Theory},
  volume  = {45},
  number  = {6},
  pages   = {1827--1832},
  year    = {1999}}

@article{HeinEisertBriegel2004,
  author  = {Hein, Marc and Eisert, Jens and Briegel, Hans J.},
  title   = {Multiparty entanglement in graph states},
  journal = {Phys. Rev. A},
  volume  = {69},
  pages   = {062311},
  year    = {2004}}

@article{Oum2005,
  author  = {Oum, Sang-il},
  title   = {Rank-width and vertex-minors},
  journal = {J. Combin. Theory Ser. B},
  volume  = {95},
  pages   = {79--100},
  year    = {2005}}

@article{BravyiTerhal2009,
  author  = {Bravyi, Sergey and Terhal, Barbara},
  title   = {A no-go theorem for a two-dimensional self-correcting quantum memory based on stabilizer codes},
  journal = {New J. Phys.},
  volume  = {11},
  pages   = {043029},
  year    = {2009}}

@article{SchlingemannWerner2001,
  author  = {Schlingemann, Dirk and Werner, Reinhard F.},
  title   = {Quantum error-correcting codes associated with graphs},
  journal = {Phys. Rev. A},
  volume  = {65},
  pages   = {012308},
  year    = {2001}}

@article{OxleyWhittle1993,
  author  = {Oxley, James and Whittle, Geoff},
  title   = {A characterization of {T}utte invariants of 2-polymatroids},
  journal = {J. Combin. Theory Ser. B},
  volume  = {59},
  pages   = {210--244},
  year    = {1993}}

@article{Greene1976,
  author  = {Greene, Curtis},
  title   = {Weight enumeration and the geometry of linear codes},
  journal = {Stud. Appl. Math.},
  volume  = {55},
  pages   = {119--128},
  year    = {1976}}

@misc{SwartzWentworthNiceXue2024,
  author = {Swartz, Ed and Wentworth-Nice, Prairie and Xue, Alexander},
  title  = {Polymatroids are to finite groups as matroids are to finite fields},
  year   = {2024},
  note   = {arXiv:2402.17582}}

@article{KnillLaflamme1997,
  author  = {Knill, Emanuel and Laflamme, Raymond},
  title   = {Theory of quantum error-correcting codes},
  journal = {Phys. Rev. A},
  volume  = {55},
  number  = {2},
  pages   = {900--911},
  year    = {1997}}

@article{BennettEtAl1993,
  author  = {Bennett, Charles H. and Brassard, Gilles and Cr{\'e}peau, Claude and Jozsa, Richard and Peres, Asher and Wootters, William K.},
  title   = {Teleporting an unknown quantum state via dual classical and {E}instein--{P}odolsky--{R}osen channels},
  journal = {Phys. Rev. Lett.},
  volume  = {70},
  number  = {13},
  pages   = {1895--1899},
  year    = {1993}}

@article{Werner2001,
  author  = {Werner, Reinhard F.},
  title   = {All teleportation and dense coding schemes},
  journal = {J. Phys. A: Math. Gen.},
  volume  = {34},
  number  = {35},
  pages   = {7081--7094},
  year    = {2001}}

@article{Bouchet1987,
  author  = {Bouchet, Andr{\'e}},
  title   = {Greedy algorithm and symmetric matroids},
  journal = {Math. Program.},
  volume  = {38},
  number  = {2},
  pages   = {147--159},
  year    = {1987}}

@article{BouchetDuchamp1991,
  author  = {Bouchet, Andr{\'e} and Duchamp, Alain},
  title   = {Representability of {$\Delta$}-matroids over {GF}(2)},
  journal = {Linear Algebra Appl.},
  volume  = {146},
  pages   = {67--78},
  year    = {1991}}

@misc{GangulyPaperI,
  author = {Ganguly, Sayantan},
  title = {Representability of Graphical Symplectic Matroids},
  year = {2026},
  note = {arXiv:2611.01241v2 [math.CO], companion paper},
  url = {https://arxiv.org/abs/2611.01241v2}}

@article{ImaiEtAl2005,
  author  = {Imai, Hideki and M{\"u}ller-Quade, J{\"o}rn and Nascimento, Anderson C. A. and Tuyls, Pim and Winter, Andreas},
  title   = {An information theoretical model for quantum secret sharing schemes},
  journal = {Quantum Inf. Comput.},
  volume  = {5},
  number  = {1},
  pages   = {69--80},
  year    = {2005}}

@article{SchumacherNielsen1996,
  author  = {Schumacher, Benjamin and Nielsen, Michael A.},
  title   = {Quantum data processing and error correction},
  journal = {Phys. Rev. A},
  volume  = {54},
  pages   = {2629},
  year    = {1996}}

@article{HilleryBuzekBerthiaume1999,
  author  = {Hillery, Mark and Bu{\v{z}}ek, Vladim{\'\i}r and Berthiaume, Andr{\'e}},
  title   = {Quantum secret sharing},
  journal = {Phys. Rev. A},
  volume  = {59},
  number  = {3},
  pages   = {1829--1834},
  year    = {1999}}

@article{NascimentoMullerQuadeImai2001,
  author  = {Nascimento, Anderson C. A. and M{\"u}ller-Quade, J{\"o}rn and Imai, Hideki},
  title   = {Improving quantum secret-sharing schemes},
  journal = {Phys. Rev. A},
  volume  = {64},
  number  = {4},
  pages   = {042311},
  year    = {2001}}

@inproceedings{ItoSaitoNishizeki1987,
  author    = {Ito, Mitsuru and Saito, Akira and Nishizeki, Takao},
  title     = {Secret sharing scheme realizing general access structure},
  booktitle = {Proc. IEEE Global Telecommunications Conf. (GLOBECOM)},
  pages     = {99--102},
  year      = {1987}}

@article{HostensDehaeneDeMoor2005,
  author  = {Hostens, Erik and Dehaene, Jeroen and De Moor, Bart},
  title   = {Stabilizer states and {C}lifford operations for systems of arbitrary dimensions and modular arithmetic},
  journal = {Phys. Rev. A},
  volume  = {71},
  number  = {4},
  pages   = {042315},
  year    = {2005}}

@article{Lovasz1965,
  author  = {Lov{\'a}sz, L{\'a}szl{\'o}},
  title   = {On graphs not containing independent circuits},
  journal = {Mat. Lapok},
  volume  = {16},
  pages   = {289--299},
  year    = {1965},
  note    = {(in Hungarian)}}

@article{Dirac1963,
  author  = {Dirac, G. A.},
  title   = {Some results concerning the structure of graphs},
  journal = {Canad. Math. Bull.},
  volume  = {6},
  pages   = {183--210},
  year    = {1963}}

@book{LidlNiederreiter1997,
  author    = {Lidl, Rudolf and Niederreiter, Harald},
  title     = {Finite Fields},
  edition   = {2nd},
  series    = {Encyclopedia of Mathematics and its Applications},
  volume    = {20},
  publisher = {Cambridge University Press},
  address   = {Cambridge},
  year      = {1997}}

@book{Oxley2011,
  author    = {Oxley, James},
  title     = {Matroid Theory},
  edition   = {Second},
  publisher = {Oxford University Press},
  address   = {Oxford},
  year      = {2011}}

@article{Gheorghiu2012,
  author = {Gheorghiu, Vlad},
  title = {Generalized semiquantum secret-sharing schemes},
  journal = {Phys. Rev. A},
  volume = {85},
  pages = {052309},
  year = {2012},
  doi = {10.1103/PhysRevA.85.052309}}

@article{GheorghiuLooiGriffiths2010,
  author = {Gheorghiu, Vlad and Looi, Shiang Yong and Griffiths, Robert B.},
  title = {Location of quantum information in additive graph codes},
  journal = {Phys. Rev. A},
  volume = {81},
  pages = {032326},
  year = {2010},
  doi = {10.1103/PhysRevA.81.032326}}

@article{Zhukov2018,
  author = {Zhukov, V. I.},
  title = {Lagrangian subspaces, delta-matroids, and four-term relations},
  journal = {Funct. Anal. Appl.},
  volume = {52},
  number = {2},
  pages = {93--100},
  year = {2018},
  doi = {10.1007/s10688-018-0215-6},
  note = {See also arXiv:1805.12477, Theorem 2.1}}

@article{Higgs1968,
  author = {Higgs, D. A.},
  title = {Strong maps of geometries},
  journal = {J. Combin. Theory},
  volume = {5},
  number = {2},
  pages = {185--191},
  year = {1968}}

@misc{Raz2024,
  author = {Raz, Or},
  title = {The geometry of ranked symplectic matroids},
  year = {2024},
  note = {arXiv:2411.09396v3 [math.CO]},
  url = {https://arxiv.org/abs/2411.09396}}
\end{document}